\documentclass[12pt,a4paper]{article}
\usepackage{amsmath,amscd}
\usepackage{amssymb}
\usepackage{bm}
\usepackage{cite}  
\usepackage{comment}
\usepackage{graphicx}
\usepackage{hyperref}
\usepackage{makeidx}
\usepackage{mathpazo} 
\usepackage{mathtools}
\usepackage{multimedia}
\usepackage{slashed}
\usepackage[small,nohug,heads=vee]{diagrams}
\usepackage[usenames,dvipsnames]{xcolor}
\usepackage{xfrac}

\definecolor{light-gray}{gray}{0.80}
\definecolor{dRed}{RGB}{0, 0, 0}
\diagramstyle[labelstyle=\scriptstyle]

\newcommand\eq{=&&\hspace{-18pt}}
\newcommand\p{\hspace{1pt}}
\newcommand\strt[1]{\rule[-#1pt]{0pt}{#1pt}}

\def\cm{\mathcal{M}}
\def\ct{\mathcal{T}}

\usepackage{fancyhdr}
 \renewcommand{\maketitle}{
     \begin{center}
       \Large
         {\bf A 4+1 Formalism for the Evolving Stueckelberg-Horwitz-Piron Metric}
         \vskip .3 true cm
       \small
         Martin Land \\
         \vskip .3 true cm
         Department of Computer Science \\
         Hadassah College \\
         37 HaNevi'im Street, Jerusalem \\
 email: martin@hac.ac.il
       \end{center}
       \vskip .5 true cm
 }
\input{tcilatex}

\begin{document}
\title{}
\author{}
\maketitle


%
\begin{abstract}
We propose a field theory for the local metric in
Stueckelberg--Horwitz--Piron (SHP) general relativity, a framework in which 
the evolution of classical four-dimensional (4D) worldlines $x^\mu \left( \tau \right)$ ($\mu =
0,1,2,3 $) is parameterized by an external time $\tau$.
Combining insights from SHP electrodynamics and the ADM formalism in general
relativity, we generalize the notion of a 4D spacetime $\mathcal{M}$ to a formal
manifold $\mathcal{M}_5 = \mathcal{M} \times R$, representing an admixture of geometry (the
diffeomorphism invariance of $\mathcal{M}$) and dynamics (the system evolution
of $\mathcal{M} \left( \tau \right)$
with the monotonic advance of $\tau \in R$).
%
Strategically breaking the formal 5D symmetry of a metric
$g_{\alpha\beta}(x,\tau)$ ($\alpha,\beta = 0,1,2,3,5 $) posed on
$\mathcal{M}_5$, we obtain ten unconstrained Einstein equations for the $\tau$-evolution of the
4D metric $\gamma_{\mu\nu}(x,\tau)$ and five constraints that are to be satisfied by the initial conditions.
The resulting theory differs from five-dimensional (5D) gravitation, much as SHP U(1) gauge theory
differs from 5D electrodynamics.
\end{abstract}

\baselineskip7mm 
\parindent=0cm \parskip=10pt

\section{Introduction}

{The Arnowitt Deser Misner (ADM)} formalism \cite{ADM} in general relativity (GR)
expresses the Einstein field equations in canonical form, thus permitting a solution of 
particular field/matter configurations formulated as initial value problems. 
As a canonical Hamiltonian formulation that splits four-dimensional (4D) spacetime into three-dimensional (3D) space and a selected time
direction, ADM provides insight into general features of relativity, but is not always the
most convenient of the 3+1 formulations for computation, especially numerical simulation.
\textcolor{dRed}{In this paper we borrow techniques from the 3+1 formalism in
order to generalize the Stueckelberg--Horwitz--Piron (SHP) theory of
classical electrodynamics \cite{Stueckelberg-1,Stueckelberg-2,HP,saad,rel-qm,RCM}
to SHP GR \cite{SHPGR,SHPGR2}.}
The~SHP \textcolor{dRed}{framework} is a covariant canonical approach to relativistic classical and quantum
mechanics, in which 4D spacetime events are defined with respect to coordinates $x^\mu$
$(\mu = 0,1,2,3)$ and an external evolution parameter $\tau$. 
Events trace out particle worldlines as functions $x^\mu (\tau)$ or $\psi (x,\tau)$ under
the monotonic advance of $\tau$, producing \textcolor{dRed}{five}
$\tau$-dependent gauge fields $a_\alpha \textcolor{dRed}{(x,\tau)}$
carrying the interaction between events. ({{
{Here and} 
throughout the SHP literature, Greek indices $\alpha,\beta, \gamma, ... , \eta $
take the values $0,1,2,3,5$, while~$\lambda, \mu, \nu, ... $ run from 0 to 3.}})
The result is an integrable electrodynamics, instantaneous in the external time $\tau$, but
recovering Maxwell theory in a $\tau$-equilibrium limit.
At numerous stages of analysis in SHP, an apparent five-dimensional (5D) symmetry arising from the five variables
$x^\mu,\tau$ must be judiciously broken to 4+1 representations of
O(3,1)\textcolor{dRed}{, because the $x^\mu$ are coordinates while $\tau$ is an external parameter.}
In this paper we apply the lessons of SHP electrodynamics to a 4+1 theory of
\textcolor{dRed}{a} local metric $g_{\alpha\beta}(x,\tau)$. 
\textcolor{dRed}{As we shall see, this approach differs from a 3+2 or (3+1)+1
formalism, in that we do not split 4D spacetime into space and time, maintaining
the manifest spacetime covariance of the underlying physical picture in each
step. 
Rather, we construct a purely formal \hbox{4+1 $\longrightarrow$
5D} manifold as a
guide to formulating field equations that under the \hbox{5D
$\longrightarrow$ 4+1} foliation describe a spacetime metric $\gamma_{\mu\nu}(x,\tau)$ evolving with $\tau$ and
preserving the required spacetime symmetries.
}

\subsection{Motivation: The Problem of Time}

In summarizing Einstein gravity as ``Spacetime tells matter how to move; matter tells
spacetime how to curve,'' Wheeler \cite{wheeler_bio} touched on certain general issues in
relativity known collectively as the problem of time
In nonrelativistic mechanics, space is viewed as the ``arena'' of physical motion, a
manifold with given background metric in some coordinate system, while time is an external
parameter introduced to mark the coordinate evolution that characterizes the motion of
objects in space. 
In contrast, time in general relativity retains its traditional Newtonian role as
evolution parameter, but also serves as a coordinate, and thus, through the metric, plays
a structural role in the spacetime ``arena'' itself.   
This dual role is complicated by the principal features of general relativity: the~diffeomorphism invariance that eliminates any \textit{
{a priori}} distinction between space
and time coordinates, and the background independence that regards gravitation as 
equivalent to motion in the spacetime determined by the local metric.

Because the metric is itself determined by the time parameterized motion of matter,
practical~approaches to problems in gravitation generally pose the Einstein field
equations and the equations of motion for matter as an initial value problem. 
Beginning with a consistent spacetime geometry at some time, one may solve for the evolution of spacetime and the motions of matter over time.  
Known as a 3+1 formalism, this approach singles out a time direction, as in standard
Hamiltonian formulations of field theory, and so the equations are not manifestly
covariant, although general covariance is preserved at each step \cite{ADM,isham,kiefer}.
On the one hand, a configuration of matter and spacetime that satisfies the equations of GR represents a 4D block universe, given once and describing all space, past, present, and future.
Additionally, on the other hand, we may find such solutions by integrating forward in time from
consistent initial conditions at some time.  
In Wheeler's words \cite{Superspace}, ``A decade and more of work by Dirac, Bergmann,
Schild, Pirani, Anderson, Higgs, Arnowitt, Deser, Misner, DeWitt, and others has taught us
through many a hard knock that Einstein's geometrodynamics deals with the dynamics of
geometry: of 3-geometry, not 4-geometry.''   

Unsurprisingly, the foliation of spacetime into 
three-geometries of simultaneous points in space further complicates the interpretation of
time. 
Because time is only felt in the evolution from one 3D submanifold to another, the
Hamiltonian is constrained to vanish when restricted to any given equal-time
three-geometry~\cite{kiefer}.  
Moreover, there is no preferred criterion for choosing a functional of canonical
variables that might be used as an intrinsic time parameter.  
While one may consider a physical clock that measures the proper time in some reference
frame, the proper time depends on a spacetime trajectory that is only known after the
equations of motion have been solved.
While~such a system may be well-posed in classical GR \cite{isham}, this is less obvious
if the metric is subject to quantum~fluctuations.  

\subsection{Stueckelberg-Horwitz-Piron (SHP) Theory}
\label{SHP}

Stueckelberg--Horwitz--Piron (SHP) theory
is a covariant approach to relativistic classical and quantum mechanics developed to
address the problem of time as it arises in electrodynamics.  
In~1937 Fock proposed using proper time as the evolution parameter for a Newton-like force
law, succinctly expressing a manifestly covariant formulation of electrodynamics \cite{Fock}. 
But, four years later, Stueckelberg proposed \cite{Stueckelberg-1,Stueckelberg-2}
to interpret antiparticles as particles moving backward in time, and~showed that neither the coordinate time $x^0 = ct$ nor the proper time of the motion could serve as
evolution parameter for particle/antiparticle pair processes. 
Because $ds^2 = \eta_{\mu\nu} dx^\mu dx^\nu$ cannot remain constant during such processes, he
introduced an external time $\tau$ and argued that $ds^2 = \eta_{\mu\nu} \dot x^\mu \dot
x^\nu d\tau^2$ can be a $\tau$-dependent dynamical quantity, even in flat space.
In 1973, Horwitz and Piron \cite{HP} were similarly led to use an external time in
formulating a manifestly covariant relativistic mechanics with interactions, in order to
overcome \textit{
{a priori}
} constraints on the 4D phase space that conflict with canonical
structure. 
Thus, writing the eight-dimensional (8D) unconstrained phase space

\vspace{-6pt}

\begin{equation}
x^\mu(\tau), \ \dot x^\mu(\tau)  \qquad \qquad  \dot x^\mu = \frac{dx^\mu}{d\tau}  
\qquad \qquad \lambda, \mu,\nu, \ldots = 0,1,2,3
\end{equation}

the O(3,1)-symmetric action for a particle in Maxwell theory 

\begin{equation}
\textcolor{dRed}{S_{\text{Maxwell}} =\int d\tau \left[ \frac{1}{2}M\dot{x}^{\mu }\dot{x}_{\mu }+\frac{e}{c}\dot{x}
^{\mu }A_{\mu }\left( x^{\lambda }\right)\right]} \mbox{\qquad}\mu ,\lambda =0,1,2,3
\label{action-1}
\end{equation}

leads to the Lorentz force in the covariant form found by Fock.  However, because the potential $A_\mu$ is produced by a Maxwell current

\begin{equation}
J^\mu (x) = \int d\tau \p \dot X^\mu (\tau) \p \delta^4 \left( x-X(\tau)\right) 
\end{equation}

depending on the trajectory $X^\mu(\tau)$ that is only given {\em after} the
equations of motion have been solved, the~system may not be well-posed.
To overcome this conflict, Horwitz, Saad, and Arshansky \cite{saad} extended the action
(\ref{action-1}) by adding $\tau $-dependence to the vector potential, along with a new scalar
potential, to~obtain the action  
\begin{eqnarray}
S_{\text{Maxwell}} \longrightarrow S_{\text{SHP}}
\eq \int d\tau ~\frac{1}{2}M\dot{x}^{\mu }\dot{x}_{\mu }+\frac{e
}{c}\dot{x}^{\mu }a_{\mu }\big( x^{\lambda },\tau \big) +\frac{e}{c}
c_{5}a_{5}\big( x^{\lambda },\tau \big)
\label{SM} \\
\eq \int d\tau ~\frac{1}{2}M\dot{x}^{\mu }\dot{x}_{\mu }
+\frac{e }{c}\dot{x}^{\beta }a_{\beta }\big( x^{\lambda },\tau \big)
\label{SSHP}
\end{eqnarray}
where $\alpha,\beta,\gamma = 0,1,2,3,5$, and in analogy to $x^0 = c t$, we write $x^5 = c_5
\tau$. 
Compatibility of SHP electrodynamics with Maxwell theory requires
\textcolor{dRed}{$c_5 \ll c$} and we will neglect $(c_5 / c)^2$ where appropriate. 
If we take the potential to be pure gauge, as $a_\alpha = \partial_\alpha
\Lambda (x,\tau)$, then the interaction term is just the total $\tau$-derivative of
$\Lambda$, showing that this theory is the most general U(1) gauge theory on the
unconstrained phase space (see also \cite{beyond}).
Variation with respect to $x^\mu$ leads to the Lorentz force \cite{lorentz} in the form   
\begin{eqnarray}
M\ddot{x}_{\mu } \eq \frac{e}{c}\left( \dot{x}^{\nu }f_{\mu \nu }+c_{5}f_{\mu
5}\right) = \frac{e}{c}\dot{x}^{\beta }f_{\mu \beta } 
\label{L-1}
\\
\frac{d}{d\tau }\left( - \frac{1}{2}M\dot{x}^{\mu }\dot{x}_{\mu }\right)
\eq  c_{5}\frac{e}{c}\dot{x}^{\beta }f_{5 \beta }
\label{L-2}
\end{eqnarray}
where the field

\begin{equation}
f_{\alpha\beta} = \partial_\alpha a_\beta - \partial_\beta a_\alpha
\end{equation}

is made a dynamical quantity by addition of a kinetic term of the type

\begin{equation}
S_{\text{field}} = \int d\tau \p d^4x f^{\alpha\beta} (x,\tau) f_{\alpha\beta} (x,\tau)
\label{f-kin}
\end{equation}
to the total action.
\textcolor{dRed}{Because the apparent 5D symmetry of the interaction term
$\dot{x}^{\beta }a_{\beta }\left( x,\tau \right)$ in the
action (\ref{SSHP}) is broken to 4+1 in (\ref{SM}), SHP
electrodynamics differs in significant ways from 5D Maxwell theory.}
We~notice that (\ref{L-2}) permits the exchange of mass between particles and fields, and
indicates the condition for non-conservation of proper time.  
It has been shown \cite{lorentz} that the total mass, energy, and~momentum of particles and fields are conserved. 

These equations of motion, along with the $\tau$-dependent field equations, have been used to calculate \cite{pair} the Bethe--Heitler mechanism for
electron-positron production in classical electrodynamics.  
A positron (an electron with $\dot x^0 = c\dot t <0$) propagates backward in coordinate time until entering the bremsstrahlung field produced by another electron scattering off a heavy
nucleus.
This field leads to $\ddot t >0$, so the particle gains energy $E = Mc^2 \p \dot
t < 0 $ continuously (and~thus $\dot x^\mu \dot x_\mu $ changes sign twice) until
emerging as an electron propagating forward in coordinate time with $E = Mc^2 \p
\dot t > 0 $.  
At coordinate times prior to the particle's turn-around (when $E = Mc^2 \p \dot t = 0$) no
particles will be observed, but two particles will be observed for subsequent coordinate times, implementing Stueckelberg's picture of pair creation.

A physical event $x^\mu (\tau)$ in SHP is an irreversible occurrence at time $\tau$ with
spacetime coordinates $x^\mu$.   
The formalism thereby implements the two aspects of time as distinct physical quantities:
the~coordinate time $x^0 = ct$ describing the locations of events, and the external
Stueckelberg time $\tau$ describing the chronological order of event occurrence. 
This eliminates grandfather paradoxes because for $\tau_2 > \tau_1$ an event $x^\mu
(\tau_2)$ at some spacetime point $x^\mu$ occurs {\em after} the event $x^\mu (\tau_1)$
and cannot affect it.   
Similarly, the 4D block universe $\mathcal{M}(\tau)$ occurs at $\tau$, representing the 4D
manifold of general relativity, comprising all of space and coordinate time $x^0$. 
A Hamiltonian $K$ generates evolution of $\mathcal{M}(\tau)$ occurring at $\tau$ to an
infinitesimally close 4D block universe $\mathcal{M}(\tau + d\tau)$ occurring at
$\tau + d\tau$. 
The~configuration of spacetime, including the past and future of $x^0 = ct$, may thus
change infinitesimally from chronological moment to moment in $\tau$.  
Thus, it is not unreasonable to expect that $\mathcal{M}(\tau)$ will be endowed with a
$\tau$-dependent metric $\gamma_{\mu\nu}(x,\tau)$ whose dynamics we explore in this paper.
On the contrary, a 4D metric given for all $\tau$ would have the character of an absolute
background field in this formalism, in violation of the goals of general relativity.

For the kinetic term (\ref{f-kin}) we formally raise the five-index of $f_{\alpha\beta}$ 
although we understand the Lagrangian density as

\begin{equation}
f^{\alpha\beta} (x,\tau) f_{\alpha\beta} (x,\tau) =
f^{\mu\nu} (x,\tau) f_{\mu\nu} (x,\tau) + 2 \sigma f^\mu_{\ 5} (x,\tau) f_{\mu 5} (x,\tau)
\end{equation}

\textcolor{dRed}{with $\sigma = \pm 1$ simply the choice of sign for the
vector-vector term.}  
That is, we bear in mind that in this notation the $\beta = 5$ index is
a formal convenience, indicating O(3,1) scalar quantities, not an
element of a 5D tensor, \textcolor{dRed}{and not a timelike coordinate.}  
In particular, $\dot x^5 = c_5$ is constrained to be a constant scalar, identical in
all reference frames, and $x^5 = c_5 \tau$ must not be treated as a dynamical variable.
Nevertheless, the contraction on indices $\alpha,\beta$ suggests a
\textcolor{dRed}{formal} 5D symmetry, possibly
O(4,1) or O(3,2) that breaks to O(3,1) in the presence of matter, and
\textcolor{dRed}{for convenience we write }

\begin{equation}
\eta_{\alpha\beta} = \text{diag} \left( -1,1,1,1,\sigma \right) 
\end{equation}

\textcolor{dRed}{in the form of} a 5D flat space metric. 
Although the higher symmetry is non-physical for matter, it~appears in wave equations, much as the wave equations for nonrelativistic acoustics appear to possess a Lorentz
symmetry not associated with the physics.  
In developing an SHP approach to general relativity, we will similarly exploit this
notation as a guide to the appropriate extension of GR while respecting the 
non-dynamical character of $x^5$.  
%

Classical and quantum SHP particle mechanics in a spacetime with
\textcolor{dRed}{a $\tau$-independent} local metric
$\gamma_{\mu\nu}(x)$ has been studied extensively by Horwitz \cite{SHPGR,SHPGR2} and will not
be discussed at length here.  
Our~goal in this paper is to find a consistent prescription for extending general
relativity to accommodate a metric $g_{\alpha\beta}(x,\tau)$ (where $\alpha,\beta =
0,1,2,3,5$) satisfying $\tau$-dependent Einstein equations on a
\textcolor{dRed}{formal} 5D manifold whose meaning
is explored through particle mechanics and field equations. 
As in standard approaches to GR, the study of embedded hypersurfaces is central to this program.
But, while the 3+1 formalism begins with a 4D block universe $\cm$ and defines a foliation
into embedded spacelike hypersurfaces of equal \textcolor{dRed}{coordinate time
$t$}, the 4+1 formalism begins with a
parameterized family of \textcolor{dRed}{4D} spacetimes $\cm(\tau)$ embedded as hypersurfaces into a 5D
pseudo-spacetime.  
Because the evolution of $\cm(\tau)$ is determined by an O(3,1) scalar Hamiltonian $K$,
with $\tau$ as an external parameter (Poincar\'e~invariant by definition), there is no
conflict with the diffeomorphism invariance of general relativity.   
This approach will guide us toward the formal structures of a 5D manifold $\cm_5$ with
coordinates $(x,\tau)$ on which we may perform a 4+1 foliation by choosing $\tau$ as the
unambiguously preferred time direction ({
{See} \cite{pitts1,pitts2} for discussion of
general 5D spacetime with preferred foliation.}).  
We refer to $\cm_5$ as a pseudo-spacetime to emphasize that despite
the formal manifold structure,
in specifying the physics we treat $\tau$ as a parameter and not a coordinate.
Moreover, $\cm_5$ represents an admixture of symmetries: 4D spacetime geometry within each
$\cm(\tau)$, and canonical dynamics between any pair $\cm(\tau_1)$, $\cm(\tau_2)$.  
We expect no general diffeomorphism invariance for $\cm_5$.

\subsection{Organization of This Paper}

The remainder of this paper is organized, as follows: in Section \ref{particle},
we formulate the particle mechanics for an event in 5D pseudo-spacetime, derive
the 5D mass-energy-momentum tensor for non-thermodynamic dust, and pose the
Einstein field equations generalized to 5D. We obtain a general solution for
the associated weak field equations, and consider a source event of slightly
varying mass (time acceleration in a co-moving frame). This leads to a small
nonrelativistic modification to Newtonian gravity in which the mass variation of
the source is transferred through the metric to induce varying mass motion in a
test event. In Section \ref{field}, we formalize the foliation of the 5D
pseudo-spacetime into the 4+1 hypersurface geometry, and by projecting onto
tangent and normal components, express~5D Einstein equations as a set of
coupled partial differential equations in the intrinsic and extrinsic curvature
of the hypersurface.
In Section \ref{ADM}, we complete the 4+1 ADM formalism by transforming the
differential equations to \textcolor{dRed}{covariant} canonical Hamiltonian
form. Finally, in Section~\ref{SG} we apply the 4+1 formalism to two possible generalizations of Schwarzschild
geometry. In the first, we~include a non-trivial fifth component in the
diagonal metric, which is seen to be constrained to satisfy a 4D wave equation.
A test event moving in the resulting field evolves with mass that depends on its
distance from the source.  In the second, we allow for the mass parameter in the
standard Schwarzschild metric to be $\tau$-dependent and find the conditions of
the mass-energy-momentum tensor that lead to such a solution.  
\textcolor{dRed}{The presented examples were chosen because they can be solved in closed form.
Realistic applications of this formalism will necessarily require numerical
solutions beyond the scope of this~paper.}

\section{Particle Mechanics}
\label{particle}
\vspace{-6pt}

\subsection{Particle Lagrangian in Standard GR}

Regarding the spacetime manifold $\cm$ as a 4D block universe, general relativity begins with consideration of the squared interval 

\begin{equation}
\delta x^2 = \gamma_{\mu\nu} \delta x^\mu \delta x^\nu = \left( x_2 - x_1 \right)^2 
\label{interval}
\end{equation}

between two neighboring points of $\cm$. 
The invariance of this interval, viewed as an instantaneous displacement in the block
universe, is a geometrical statement referring to the freedom that is permitted in
assigning a coordinate map to the manifold. 
To extract dynamics from geometry, one considers the spacetime trajectory of a material
event ({
{some appropriate} abstraction of point mass, which in GR
would necessarily be a black hole}), described as a mapping of an arbitrary parameter
$\zeta$ to a continuous sequence of events $x^\mu(\zeta)$ in $\cm$.   
Because the interval between any two points on a trajectory must be timelike, the proper
time $s$ may be taken as parameter, and ``motion'' along the trajectory is observed
through advances in the time coordinate $x^0(s)$ for advancing values of $s$.  
The invariant interval (\ref{interval}) can be written

\begin{equation}
\delta x^2 = \gamma_{\mu\nu} \delta x^\mu \delta x^\nu
= \gamma_{\mu\nu} \frac{dx^\mu}{ds}\frac{dx^\nu}{ds}  \delta s^2
= \gamma_{\mu\nu} \dot x^\mu \dot x^\nu  \delta s^2 
\label{interval-2}
\end{equation}

suggesting \cite{DiracGR} a dynamical description of the trajectory by the action

\begin{equation}
S = \int dx  = \int ds \ \sqrt{-\gamma_{\mu\nu} \dot x^\mu \dot x^\nu }
\label{sqrt}
\end{equation}

and leading to geodesic equations of motion as an expression of the equivalence
principle.  
The geodesic equations can also be derived from the action

\begin{equation}
S = \int ds \ \frac{1}{2} \ \gamma_{\mu\nu} \dot x^\mu \dot x^\nu 
\end{equation}

which removes the constraint $\dot x^2 = -c^2$ associated with (\ref{sqrt}).

\subsection{Particle Lagrangian in SHP GR}

To extend the SHP classical mechanics of a free particle to a manifold with a
$\tau$-dependent local metric, we begin by considering the
interval
\begin{equation}
dx^\mu =  x^\mu_1 (\tau_1 ) - x^\mu_2 (\tau_2)
\end{equation}
between an event $x^\mu_1 \in \cm(\tau_1)$ and an event $ x ^\mu_2 \in \cm(\tau_2) $.
Writing these events as

\begin{equation}
X_1 = (x_1,c_5 \tau_1) \qquad \qquad X_2 = ( x_2 , c_5  \tau_2)
\end{equation}

we introduce a notion of 5D distance by combining the {\em geometrical distance} $\delta x$
between any two arbitrary points 
in $\cm(\tau)$, with the {\em dynamical distance} between events generated by a Hamiltonian
that evolves $\cm(\tau) \longrightarrow \cm(\tau + \delta \tau)$. 
The geometrical distance is characterized by the squared relativistic interval
(\ref{interval}) and taking $ \tau_2 = \tau_1 + \delta \tau$, so that  

\begin{equation}
x_2 (\tau_1 + \delta \tau) - x_1(\tau_1) \simeq  x_2 (\tau_1)+ \frac{dx (\tau)}{d\tau}
\delta \tau - x_1(\tau_1) =
\delta x  + \frac{dx (\tau)}{d\tau}   \delta \tau
\end{equation}

and we write the difference in the form

\begin{equation}
X_2 - X_1 = \left( \delta x  + \frac{d{ x} (\tau)}{d\tau} \delta \tau ,
c_5 \delta \tau \right)  
\label{difference}
\end{equation}

which motivates the notion of a 5D invariant interval through


%
\begin{equation}
dX^2 =  \gamma_{\mu\nu} \left( \delta x^\mu  + \frac{d{ x}^\mu (\tau)}{d\tau}   \delta \tau
\right) \left( \delta x^\nu  + \frac{d{ x}^\nu (\tau)}{d\tau}   \delta
\tau  \right) 
+ \sigma  c_5^2 \delta \tau^2
= g_{\alpha\beta} \left( x,\tau\right)  \delta x^\alpha \delta x^\beta
\label{5D-interval}
\end{equation}

referred to $x_1$ coordinates at $\tau =\tau_1$.
\textcolor{dRed}{Because the manifold $\cm(\tau) $ evolves, the spacetime metric
$\gamma_{\mu\nu}$ must depend on $x$ and $\tau$ in some manner to be determined.}


%
%

As in 4D general relativity, the squared interval (\ref{5D-interval}) suggests the
Lagrangian

\begin{equation}
L=\frac{1}{2}Mg_{\alpha\beta}\big( x^\mu,x^5\big)  \dot{x}^{\alpha}\dot{x}^{\beta}
\qquad \lambda,\mu,\nu=0,1,2,3
\qquad \alpha,\beta,\gamma=0,1,2,3,5
\label{5-lag}
\end{equation}

from which we may find equations of motion in the space determined by the local metric 
$g_{\alpha\beta}$.

\subsection{Equations of Motion}

\textcolor{dRed}{Before examining particle dynamics in SHP GR, we consider
a straightforward extension of GR to unbroken 5D, with coordinates $x^\alpha$, for $\alpha
= 0,1,2,3,5$ and external evolution parameter $\tau$.
Naively~applying the Euler--Lagrange equations to the action (\ref{5-lag}), 
posing no fixed relationship between $x^5$ and $\tau$, we find 
}

\begin{equation}
0  = \frac{d}{d\tau }\frac{\partial L}{\partial \dot{x}^{\gamma}}-\frac{\partial L
}{\partial x^{\gamma}}
 = \frac{d}{d\tau }\left( g_{\alpha\gamma}\dot{x}^{\alpha}\right) -\frac{1}{2}\frac{
\partial }{\partial x^{\gamma}}g_{\alpha\beta}\dot{x}^{\alpha}\dot{x}^{\beta}
\end{equation}

leading to the \textcolor{dRed}{five} geodesic equations

\begin{equation}
0 =\frac{D\dot{x}^{\gamma}}{D\tau}=\ddot{x}^{\gamma}+\Gamma _{\alpha\beta}^{\gamma}\dot{x}^{\alpha}\dot{x} ^{\beta}
\label{motion-1}
\end{equation}
where $D/D\tau$ is the absolute derivative (in the notation of Weinberg
\cite{Weinberg}) and 

\begin{equation}
\Gamma _{\alpha \beta}^{\gamma }=g^{\gamma \delta}\Gamma _{\delta \alpha
\beta}=\frac{1}{2}g^{\gamma \delta}\left( \partial _{\alpha }g_{\delta
\beta}+\partial _{\beta}g_{\delta \alpha}-\partial _{\delta
}g_{\beta\alpha}\right) 
\label{connection}
\end{equation}

is the standard Christoffel symbol in 5D.  Writing the canonical momentum 

\begin{equation}
p_{\alpha} = \frac{\partial L}{\partial \dot{x}^{\alpha}}=Mg_{\alpha\beta}\dot{x}^{\beta} \qquad
\longrightarrow \qquad 
\dot{x}^{\alpha}  = \frac{1}{M}g^{\alpha\beta}p_{\beta}
\end{equation}

the Hamiltonian 

\begin{equation}
K=\dot{x}^{\alpha}p_{\alpha}-L = \frac{1}{2M}g^{\alpha\beta}p_{\alpha}p_{\beta} = L
\end{equation}

is conserved, as seen directly through

\begin{equation}
\frac{d}{d\tau}\left(
\frac{1}{2}Mg_{\alpha\beta}\dot{x}^{\alpha}\dot{x}^{\beta}\right) 
=Mg_{\alpha\beta}\dot{x}^{\alpha}\frac{D\dot{x}^{\beta}}{D\tau}=0
\end{equation}

where we used metric compatibility

\begin{equation}
\frac{Dg_{\alpha\beta}}{D\tau}=0 .
\end{equation}

Time independence of the Hamiltonian may also
be found from the canonical equations of motion 

\begin{equation}
\dot{x}^{\alpha}=\frac{dx^{\alpha}}{d\tau }=\frac{\partial K}{\partial p_{\alpha}}\qquad
\qquad \dot{p}_{\alpha}=\frac{dp_{\alpha}}{d\tau }=-\frac{\partial K}{\partial x^{\alpha}}
\end{equation}

and the Poisson bracket

\begin{equation}
\left\{ F,G\right\} =\frac{\partial F}{\partial x^{\alpha}}\frac{\partial G}{
\partial p_{\alpha}}-\frac{\partial F}{\partial p_{\alpha}}\frac{\partial G}{\partial
x^{\alpha}} 
%
\end{equation}

so that

\begin{equation}
\frac{d}{d\tau}\left( \frac{1}{2M}g^{\alpha\beta}p_{\alpha}p_{\beta} \right)
=\frac{dK}{d\tau}=\left\{ K,K\right\} +\frac{\partial K}{\partial \tau}=\frac{1
}{2M}  p_{\alpha}p_{\beta} \ \frac{\partial g^{\alpha\beta}}{\partial \tau} =0 
\end{equation}

because the metric is not explicitly dependent on $\tau$, which
\textcolor{dRed}{in this case} bears no specific
relationship with $x^5$.  As seen in SHP
electrodynamics, the equation

\begin{equation}
0 =\frac{D\dot{x}^{5}}{D\tau}=
\ddot{x}^{5}+\Gamma _{\alpha\beta}^{5}\dot{x}^{\alpha}\dot{x} ^{\beta}
\label{unphysical}
\end{equation}

\textcolor{dRed}{cannot generally be made consistent with the SHP condition} 
$x^5 =c_5 \tau \ \Rightarrow \ \ddot{x}^{5} = 0$.
Rather, \textcolor{dRed}{the SHP formalism defines $x^5 $ to be a scalar, in
which case the absolute derivative reduces to the total derivative, so that
}

\begin{equation}
\frac{D\dot{x}^{5}}{D\tau}= \frac{d\dot{x}^{5}}{d\tau}= 0
\label{physical}
\end{equation}

\textcolor{dRed}{will replace} (\ref{unphysical}).

To obtain the correct equations of motion \textcolor{dRed}{for SHP},
we must break the 5D symmetry of (\ref{5-lag})
to 4+1 prior to applying the Euler--Lagrange equations and not treat $x^5$ as a dynamical
quantity.
Expanding

\begin{equation}
L=\dfrac{1}{2}Mg_{\alpha\beta}(x,\tau)\dot{x}^{\alpha}\dot{x}^{\beta}
= \dfrac{1}{2}Mg_{\mu\nu}\; \dot{x}^{\mu}\dot{x}^{\nu}
+ Mc_5 \; g_{\mu 5}\dot{x}^{\mu}
+ \dfrac{1}{2}Mc^2_5 \; g_{55}
\label{EL-2}
\end{equation}

the equations of motion have four components

\begin{equation}
\ddot{x}^{\mu}+\Gamma _{\lambda \sigma
}^{\mu }\dot x^\lambda \dot x^\sigma +2c_5\Gamma _{5\sigma
}^{\mu }\dot x^\sigma +c^2_5\Gamma _{55}^{\mu } = 0
\label{eq-m}
\end{equation}

and, because $x^5$ is not a dynamical quantity, it has no conjugate momentum. 
Thus, while (\ref{eq-m}) is identical to (\ref{motion-1}) for \textcolor{dRed}{$\mu =
0,1,2,3$}, we understand (\ref{unphysical}) in the sense of (\ref{physical}).
\textcolor{dRed}{The breaking of 5D symmetry is expressed here in that $\Gamma
_{\alpha\beta}^{5}$ can be calculated, but it plays no part in the equations of
motion.}
The~4-momentum is

\vspace{-6pt}

\begin{equation}
p_{\mu} = \frac{\partial L}{\partial \dot{x}^{\mu}}=Mg_{\mu\nu}\dot{x}^{\nu} + Mc_5 \; g_{\mu 5}
\qquad \longrightarrow \qquad \dot{x}_{\mu} 
= \frac{1}{M} \left( p_{\mu} - Mc_5 \; g_{\mu 5}\right) 
\label{b-mom}
\end{equation}

allowing us to write the Hamiltonian in the form

\begin{equation}
K  = p_{\mu }\dot{x}^\mu -L
= \left( Mg_{\mu\nu}\dot{x}^{\nu} + Mc_5 \; g_{\mu 5} \right) \dot{x}^\mu -L
= \frac{1}{2}Mg_{\mu\nu}\dot{x}^{\mu}\dot{x}^{\nu} - \frac{1}{2}Mc^2_5 g_{55}
\label{Ham-1}
\end{equation}

which, unlike the Hamiltonian for unbroken 5D symmetry, is not equal to the
Lagrangian ({
{The~difference} is precisely the term $p_5 \dot x^5$ that would be
present in the Legendre transformation if we had taken $x^5$ to be dynamical}).
Taking the total $\tau$-derivative of (\ref{Ham-1}) and inserting the equations of motion
(\ref{eq-m}) leads to 

\begin{equation}
\frac{d K}{d \tau }  = 
-\frac{1}{2}M\dot{x}^{\mu }\dot{x}^{\nu }\frac{\partial g_{\mu
\nu }}{\partial \tau }-\frac{1}{2}Mc_{5}^{2}\frac{\partial g_{55}}{
\partial \tau }
\label{non-cons}
\end{equation}

showing that this Hamiltonian is not conserved for a $\tau$-dependent metric.
Using (\ref{b-mom}) to eliminate $\dot{x}_{\mu}$, we put the Hamiltonian into
the form 

\begin{equation}
K = \frac{1}{2M}g^{\mu\nu}p_{\mu}p_{\nu} - c_5 g^\mu_5 p_\mu + \frac{1}{2}M c_5^2 
\left( g^\mu_{\ 5} g_{\mu 5} - g_{55} \right) 
\label{Ham-2}
\end{equation}
and find its non-conservation from the Poisson bracket 

\begin{equation}
\frac{d K}{d \tau }  = \{K,K\} +
\frac{\partial K}{\partial \tau }  =  - \frac{1}{2M}p^{\mu}p^{\nu}
\frac{\partial g_{\mu\nu}}{\partial \tau} - c_5 p_\mu \frac{\partial g_{\mu 5}}{\partial \tau}
+ \frac{1}{2}M c_5^2  \left( 2 g^\mu_{\ 5} \frac{\partial g_{\mu 5}}{\partial \tau}
- \frac{\partial g_{55}}{\partial \tau} \right)
\label{non-cons-2}
\end{equation}

where we used 

\begin{equation}
\frac{ \partial g^{\mu \nu }}{\partial \tau } = - g^{\mu \rho }g^{\nu \sigma }\frac{\partial
g_{\rho \sigma }}{\partial \tau } \ . 
\end{equation}

{
{When} 
 $g_{\alpha 5} = 0$, the Hamiltonian (\ref{Ham-2}) is
seen to generalize the nonrelativistic expression ${\mathbf p}^2/2m$ for the
energy of a free particle. Because $K$ is a Lorentz scalar, SHP theory associates this
Hamiltonian with the dynamical mass of the particle motion.} 
Section \ref{weak} provides an example of a test particle evolving with variable mass in a $\tau$-dependent local metric.

\subsection{Mass-Energy-Momentum Tensor}

When considering non-thermodynamic dust, we define $n(x,\tau)$ to be the number of events
per spacetime volume, 
and  

\begin{equation}
j^{\alpha }\left( x,\tau \right) =\rho(x,\tau) \dot{x}^{\alpha }(\tau) =M
n(x,\tau)\dot{x}^{\alpha }(\tau)
\end{equation}

is the five-component event current.  The continuity equation in flat space is  

\begin{equation}
\partial_\alpha j^\alpha = \partial_\mu j^\mu  + \partial_5 j^5 
= \partial_\mu j^\mu  + \frac{\partial \rho}{\partial \tau} =0
\end{equation}

and with a local metric is generalized to

\begin{equation}
\nabla _{\alpha }j^{\alpha } =0
\end{equation}

where (in the notation of Wald \cite{Wald}), the covariant derivative for a vector is

\begin{equation}
\nabla_\alpha X^\beta = \frac{\partial X^\beta}{\partial x^\alpha} +
X^\gamma \Gamma^\beta_{\gamma\alpha} \ .
\label{cov-div}
\end{equation}

\textcolor{dRed}{
{But again,} since $j^5$ is a scalar (the number density is scalar on physical
grounds) for which the covariant derivative is just the partial derivative, we must have }

\begin{equation}
\nabla_5 j^5 = \frac{\partial \rho}{\partial \tau}
\label{no-g5-2}
\end{equation}

so the continuity equation becomes 
\begin{equation}
\frac{\partial \rho}{\partial \tau} + \nabla _{\mu }j^{\mu } =0 . 
\end{equation}
%


Generalizing the 4D stress-energy-momentum tensor to 5D, we write the
mass-energy-momentum tensor \cite{antenna} as

\begin{equation}
T^{\alpha \beta }=
\rho \dot{x}^{\alpha } \dot{x}^{\beta } \ \longrightarrow \ \left\{
\begin{array}{l}
T^{\mu \nu }=\rho \dot{x}^{\mu } \dot{x}^{\nu } \strt{8} \\ 
T^{5\beta} =c_{5}j^\beta
\end{array}
\right.
\end{equation}

where, in addition to the 4D components $T^{\mu\nu}$, we have the current density $
T^{5\beta}=\dot{x}^{5}\dot{x}^{\beta}\rho =c_{5}j^\beta$.
The~conservation equation is 

\begin{equation}
0= \nabla_\beta T^{\alpha \beta }=\nabla_\beta \left( \rho \dot{x}^{\alpha }
\dot{x}^{\beta }\right)
=\dot{x}^{\alpha }\nabla_\beta \left( \rho 
\dot{x}^{\beta }\right)+\rho \dot{x}^{\beta }\nabla_\beta \dot{x}^{\alpha }
=\dot{x}^{\alpha }\nabla_\beta j^{\beta }
+\rho \dot{x}^{\beta }\nabla_\beta \dot{x}^{\alpha }
\end{equation}

which vanishes by virtue of the continuity and geodesic equations

\begin{equation}
\nabla _{\alpha }j^{\alpha } =0 \qquad \qquad 
\dot{x}^{\beta }\nabla_\beta \dot{x}^{\alpha }=\frac{D\dot{x}
^{\alpha }}{D\tau }=0 
\end{equation}

when the equations of motion (\ref{motion-1}) are
evaluated in the sense of (\ref{physical}).   

\subsection{Weak Field Approximation}
\label{weak}

\textcolor{dRed}{As a first step in obtaining field equations for $g_{\alpha
\beta }$ } we extend
the Einstein equations to 5D as 

\begin{equation}
G_{\alpha \beta} =
R_{\alpha \beta} - \frac{1}{2} Rg_{\alpha \beta} = \frac{8\pi G}{c^4} T_{\alpha \beta}
\end{equation}

where the Ricci tensor $R_{\alpha \beta}$ and scalar $R$ are obtained by contracting
indices of the 5D curvature tensor $R_{\gamma \alpha \beta}^\delta$.
The weak field approximation \textcolor{dRed}{(see for example
\cite{MTW,DiracGR,AS})} 
is generalized to \textcolor{dRed}{SHP GR} by introducing a
perturbation $h_{\alpha \beta }$ to the flat metric, such that

\begin{equation}
%
%
g_{\alpha \beta }=\eta _{\alpha \beta }+h_{\alpha \beta } \ \longrightarrow \ \partial_\gamma g_{\alpha\beta}
= \partial_\gamma h_{\alpha\beta}  \qquad \qquad \left( h_{\alpha \beta }\right) ^{2}\approx 0
\end{equation}

leading to the Ricci tensor

\begin{equation}
R_{\alpha \beta }\simeq \frac{1}{2}\left( 
\partial_{\beta }\partial_{\gamma }h_{\alpha }^{\gamma } + \partial_{\alpha
}\partial_{\gamma }h_{\beta }^{\gamma }- \partial_{\gamma }\partial_{\gamma
}h_{\alpha \beta } - \partial_{\alpha }\partial_{\beta }h  
\right)
\qquad R\simeq \eta ^{\alpha \beta }R_{\alpha \beta }
\qquad h\simeq\eta ^{\alpha \beta }h_{\alpha \beta } 
\end{equation}

which naturally contains only the perturbation.
Defining $\bar{h}_{\alpha \beta }=h_{\alpha \beta }-\frac{1}{2}\eta _{\alpha
\beta }h$, the Einstein equations~become

\begin{equation}
\frac{16\pi G}{c^{4}}T_{\alpha \beta }=
\partial_{\beta }\partial_{\gamma }\bar{h}_{\alpha }^{\gamma } + \partial_{\alpha
}\partial_{\gamma }\bar{h}_{\beta }^{\gamma }- \partial_{\gamma }\partial_{\gamma
}\bar{h}_{\alpha \beta } - \partial_{\alpha }\partial_{\beta }\bar{h}
\end{equation}

which take the form of a wave equation

\begin{equation}
\frac{16\pi G}{c^{4}}T_{\alpha \beta }
=-\partial ^{\gamma }\partial _{\gamma }\bar{h}_{\alpha \beta }
=-\left( \partial ^{\mu }\partial _{\mu } + \frac{\eta_{55}}{c_5^2}
\partial_\tau^2 \right) \bar{h}_{\alpha \beta }
\end{equation}

by imposing the usual gauge condition
$\partial_\lambda \bar{h}^{\alpha \lambda } =0$.  
The principal part Green's function \cite{green} for this wave equation is
\begin{equation}
G(x,\tau ) = -{\frac{1}{{2\pi }}}\delta (x^{2})\delta (\tau )-{\frac{c_5}{{
2\pi ^{2}}}}{\frac{\partial }{{\partial {x^{2}}}}}{\theta (-\eta_{55}g_{\alpha
\beta }x^{\alpha }x^{\beta })}{\frac{1}{\sqrt{-\eta_{55}g_{\alpha \beta
}x^{\alpha }x^{\beta }}}}
\end{equation}

in which the first term is dominant at long distance, leading to the solution

\begin{equation}
\bar{h}_{\alpha \beta }\left( x,\tau \right)
=\frac{4G}{c^{4}}\int d^{3}x^{\prime }\frac{T_{\alpha \beta }\left( t-
\frac{\left\vert \mathbf{x}-\mathbf{x}^{\prime }\right\vert }{c},\mathbf{x}
^{\prime },\tau \right) }{\left\vert \mathbf{x}-\mathbf{x}^{\prime
}\right\vert }  
\end{equation}

relating 
the field $\bar{h}_{\alpha \beta }\left( x,\tau \right)$ to the
source $T_{\alpha \beta }\left( x,\tau \right)$. 

\textcolor{dRed}{As a simple example, }
we consider a source $X = (cT(\tau),{\mathbf 0})$ in a co-moving
frame, so that $\dot T \ne $ constant corresponds to a variation in energy without
corresponding variation in momentum, producing a variation in mass.
The non-zero components of the mass-energy-momentum tensor are

\begin{equation}
T^{00}  = mc^{2}\dot{T}^{2}\delta ^{3}\left( \mathbf{x} \right)
\rho \left( t-T\left( \tau \right) \right)
\qquad  \
T^{\alpha i} = 0
\qquad  \
T^{55} = \frac{c_5^2}{c^2} T^{00} \approx 0
\end{equation}

where we neglect $c_5^2 / c^2 \ll 1$ and have written $M(\tau) 
= m \, \rho \left( t-T\left( \tau \right) \right)$ to represent a slowly varying density
function \textcolor{dRed}{(the source is sharply located in space but smeared
along the $t$-axis)}.
The perturbed metric is found to be

\begin{equation}
\bar{h}^{00}\left( x,\tau \right) = \frac{4GM}{c^{2}R} \dot T^2 
\qquad 
\bar{h}^{\alpha i}\left( x,\tau \right) = 0
\qquad 
\bar{h}^{55}\left( x,\tau \right) = 0
\label{pert-1}
\end{equation}

so using  
$h_{\alpha \beta }=\bar{h}_{\alpha \beta }-\frac{1}{2}\eta _{\alpha \beta }
\bar{h}$, we see that $h^{00}=\bar{h}^{00}$.
Since $ g^{\alpha\beta} h_{\beta\gamma} \simeq \eta^{\alpha\beta} h_{\beta\gamma}$
the non-zero Christoffel symbols are

\begin{equation}
\Gamma _{00}^{\mu } = -\dfrac{1}{2}\eta ^{\mu \nu }\partial _{\nu }h_{00}
\qquad \qquad 
\Gamma _{0i}^{\mu } =\dfrac{1}{ 2}\eta ^{\mu \nu }\partial _{i}h_{\nu 0}
\qquad \qquad 
\Gamma _{50}^{\mu } =\dfrac{1}{2c_{5}}\eta ^{\mu 0}\partial _{\tau }h_{00}
\end{equation}

and the equations of motion for a distant test particle split into

\begin{equation}
\ddot{t} = \left( \partial _{\tau }h_{00}\right) \dot{t}+
\mathbf{\dot{x}}\cdot \left( \nabla h_{00}\right) \dot{t}^{2}
\qquad \qquad 
\mathbf{\ddot{x}} = \frac{c^{2}}{2}\left( \nabla h_{00} \right) 
\dot{t}^{2}  
\end{equation}

where the factor $ \partial _{\tau }h_{00}$ distinguishes these equations
from the Newtonian model.
We write the space part in spherical coordinates, putting $\theta =\pi /2$, so
that the angular and radial equations become

\begin{equation}
2\dot{R}\dot{\phi}+R\ddot{\phi}=0 \ \longrightarrow \ \dot{\phi}=\frac{L}{MR^{2}}
\ \longrightarrow \ 
\ddot{R}-\frac{L^{2}}{M^{2}R^{3}}=-\frac{GM}{R^{2}}\dot{t}^{2}\dot{T}^{2} 
\end{equation}

\textcolor{dRed}{where $L$ is a constant of integration with units of angular
momentum. Introducing $\alpha \left( \tau \right)$ through} 

\begin{equation}
\dot{T}  = 1+\frac{\alpha \left( \tau \right)}{2} \ \longrightarrow \ \dot{T}^{2}\simeq
1+\alpha \left( \tau \right)  \ \longrightarrow \  \dot{T}\ddot{T} \simeq
\left(  1+\frac{\alpha \left( \tau \right)}{2}\right) \frac{\dot{
\alpha}\left( \tau \right)}{2} 
\end{equation}

\textcolor{dRed}{the relationship between $t$ and $\tau$} becomes

\begin{equation}
\ddot{t}=\frac{2G\partial _{\tau }M}{c^{2}R}\dot{t}+\frac{4GM}{c^{2}R}\dot{T}
\ddot{T}\dot{t}-\frac{2GM}{R^{2}c^{2}}\dot{R}\dot{T}^{2} 
\approx  \frac{2GM}{c^{2}R} \left(  1+\frac{\alpha \left( \tau \right)}{2}\right)
\dot{ \alpha}\left( \tau \right)\dot{t}
\end{equation}

where we neglect the nonrelativistic velocity $\dot R /c \approx 0$ 
and the slow variation in the source distribution $\partial_\tau \rho \approx 0$.
In the \textcolor{dRed}{absence of the mass perturbation, we have}
$\alpha = 0 \longrightarrow \dot t = 1$, \textcolor{dRed}{recovering a Newtonian
notion of time,} but this $t$ equation has the solution

\begin{equation}
\dot{t} = \exp \left[ \frac{2GM}{c^{2}R}\left( \alpha 
+\frac{1}{4}\alpha ^{2} \right) \right]
\ \longrightarrow \ \textcolor{dRed}{\dot{t}^{2}\dot{T}^{2} 
\simeq 1+\left( 1+\frac{4GM}{c^{2}R}\right) \alpha }
\end{equation}

\textcolor{dRed}{indicating a more complicated relationship between $t$ and $\tau$.  
Since $4GM / c^2 R \ll 1$,} this leads finally to a radial equation in the form

\begin{equation}
\textcolor{dRed}{\frac{d}{d\tau }\left\{
\frac{1}{2}\dot{R}^{2}+\frac{1}{2}\frac{L^{2}}{
M^{2}R^{2}}-\frac{GM}{R}\left[ 1+\alpha \left( \tau \right) \strt{5}
\right] \right\} = \frac{dK}{d\tau} =-\frac{GM}{R}\frac{d}{d\tau }\alpha \left( \tau \right) .}
\end{equation}
%

{We} recognize $K$ on the LHS as the Hamiltonian of the test particle moving in this
local metric, recovering~the Newtonian expression when the perturbation
$\alpha(\tau) $ vanishes. 
The mass fluctuation of the point source is seen to induce a fluctuation in the mass of
the distant test particle, acting through the field $g_{\alpha\beta} (x,\tau)$ in order to produce
a small modification of Newtonian gravity. 

\section{Field Equations}
\label{field}

In a 3+1 formalism such as ADM, a spacetime trajectory is defined with respect to a
foliation of $\cm$.   
For any point $x^\mu \in \cm$, we define a time function $t(x)$ on $\cm$ whose level sets

\begin{equation}
\Sigma (t_0) = \left\{ x^{\mu } \ \big\vert \  t(x) = t_0  \right\}  
\label{E-255}
\end{equation}

are hypersurfaces of constant time. 
The 4D hypersurface $\Sigma (t_0) \subset \cm $ is homeomorphic to a spacelike 3D
submanifold $\hat \Sigma$ with coordinates $x^i$, $i=1,2,3$, and the homeomorphism forms
an embedding of $\hat \Sigma$ into $\cm$, which may be expressed as  

\begin{equation}
x^\mu_{t_0} = x^\mu ({\mathbf x},t_0) 
\end{equation}

for fixed $t_0$.
The trajectory 

\begin{equation}
x_{{\mathbf x}_0}^{\mu }\left( t \right) = x^{\mu }\left( {\mathbf x}_{0},t \right)
\end{equation}

associated with this embedding connects the point ${\mathbf x}_0$ with fixed 3D
coordinates on different hyperspaces, suggesting a notion of time evolution from one
hyperspace to the next.  

We extend these ideas to SHP general relativity, taking advantage of the analogy with
the 3+1 formalism \textcolor{dRed}{\cite{ADM,isham,Bertschinger,zilhao}} 
and employing its standard notation.  
Roughly following the tutorial exposition of 3+1 numerical relativity that is given in
\cite{Gourgoulhon,Blau}, \textcolor{dRed}{we} decompose the Einstein field equations
into spacetime and $\tau$ sectors, leading to a set of coupled partial
differential equations for the phase space variables of the field theory,
$\gamma_{\mu\nu} ( x,\tau ) $ and $ \dot \gamma_{\mu\nu} ( x,\tau ) = \partial
\gamma_{\mu\nu} ( x,\tau ) / \partial\tau$. 
\textcolor{dRed}{Although the general presentation is familiar, it differs in
certain details, because the foliation is natural and the field theory is presumed to carry the
factor $\sigma$ associated with objects carrying a five-index.}
With appropriate initial conditions for the metric and the matter distribution, this 
poses an initial value problem that can be integrated forward in $\tau$ to solve for
evolving spacetime configurations.

\subsection{Embedding and Foliation}
\label{embedding}

The first step is to introduce a 5D pseudo-spacetime by defining the injective mapping

\begin{equation}
\Phi: \mathcal{M} \ \longrightarrow \ \mathcal{M}_5 = \mathcal{M} \times R
\qquad \qquad \qquad X = \Phi(x,\tau) = (x,c_5\tau)
\end{equation}
with coordinates $X^\alpha \in \mathcal{M}_5 $, for $\alpha=0,1,2,3,5$.
This structure admits the natural foliation defined by level surfaces of the
scalar field $\tau(X) = \tau$ 

\begin{equation}
\Sigma (\tau_0) = \left\{ X \in \cm_5 \ \big\vert \ \tau(X) = X^5 / c_5 = \tau_0  \right\} 
\end{equation}

which is homeomorphic to $\cm(\tau_0)$ for any $\tau_0$ (and so we drop reference to
$\tau_0$ in referring to the hypersurfaces).
We take

\begin{equation}
E_{\mu }^{\alpha } = \left( \frac{ \partial X^{\alpha }\left( x,\tau \right) }{\partial
x^{\mu }}\right) _{\tau _{0}}   
\qquad \qquad \mu = 0,1,2,3
\label{E-6}
\end{equation}

as the four basis elements $E_\mu = \partial_\mu $ for $\ct\left( \Sigma \right) $,
the tangent space of $\Sigma $.  Thus, when restricted to $X \in \Sigma $, the squared interval becomes 

\begin{equation}
\left. dX^2 \right\vert_{\Sigma} = \left. g_{\alpha \beta }dX^{\alpha }dX^{\beta
}\right|_{\Sigma } = g_{\alpha \beta } \frac{\partial X^{\alpha }
}{\partial x^{\mu }}\frac{\partial X^{\beta }}{\partial x^{\nu }}dx^{\mu }dx^{\nu }
= \gamma_{\mu \nu } dx^{\mu }dx^{\nu }  
\label{E-14}
\end{equation}

where we identify $ \gamma _{\mu \nu } = g_{\alpha \beta }E_{\mu }^{\alpha }E_{\nu
}^{\beta }$, the induced metric on $\Sigma  $, 
%
%
with the 4D spacetime metric we began with.
For a vector in the time direction of $\ct(\cm_5) $, we write

\begin{equation}
\partial_\alpha \p \tau(X) = \delta_\alpha^5 \p \partial_5 \p \tau(X)
= \delta_\alpha^5 \p \frac{1}{c_5} \p \partial_\tau \p \tau(X)
\label{E-268}
\end{equation}

which is normal to the tangent space of $ \Sigma $ in the sense that $\tau
\left( X \right) = \tau_0  $ is constant throughout $\Sigma (\tau_0)$. 
Thus, in $\ct(\cm_5) $, the vector $(E_5)_\alpha = \partial_\alpha \p \tau(X)$ points out
of $\ct\left( \Sigma  \right) $ in the direction of time evolution.   
The~unit normal $n_\alpha$ in the time direction is defined as

\begin{equation}
n = \sigma \frac{1}{\sqrt{\left\vert g^{55}\right\vert}} E_5 \ \longrightarrow \
n^2 = \frac{1}{\left\vert g^{55}\right\vert}
g^{\alpha \beta} (E_5)_\alpha(E_5)_\beta 
= \frac{1}{\left\vert g^{55}\right\vert} g^{55} = \sigma 
\end{equation}

so that

\begin{equation}
n^\alpha = g ^{\alpha \beta }n_{\beta }=g ^{\alpha \beta }\sigma 
\frac{1}{\sqrt{\left\vert g ^{55}\right\vert }}\delta _{\beta
}^{5}=\sigma g ^{\alpha 5}\frac{1}{\sqrt{\left\vert g
^{55}\right\vert }}  \ . 
\label{E-44}
\end{equation}
%

{For} any vector $A \in \ct \left( \cm_5 \right) $ in the tangent space of $\cm_5$ we can
project onto parallel and normal~components 

\begin{equation}
A_{\parallel } = \sigma \left( A\cdot n \right) n  \qquad \qquad \qquad 
A_{\perp } = A-\sigma \left( A\cdot n \right) n 
\end{equation}

and so define the normal projection operator
\begin{equation}
\Pi_{\alpha \beta }= \sigma n_{\alpha }n_{\beta } \qquad \qquad \qquad 
\Pi_{\alpha \gamma }\Pi^{\gamma \beta }
= \sigma^2 n^2 \; n_{\alpha }n^{\beta }= \Pi_{\alpha }^{\beta }  
\end{equation}

and the tangent projection operator

\begin{equation}
P_{\alpha \beta }=g_{\alpha \beta }-\sigma n_{\alpha }n_{\beta } \qquad P^{\alpha\beta}
=g^{\alpha\beta} -\sigma n^{\alpha }n^{\beta } \qquad P_{\alpha \gamma }P^{\gamma \beta
}=P_{\alpha }^{\beta } = \delta_\alpha^\beta -\sigma n_{\alpha }n^{\beta } 
\label{projector}
\end{equation}

along with the completeness relation

\begin{equation}
g_{\alpha\beta} = P_{\alpha \beta } + \sigma n_{\alpha }n_{\beta }
\qquad \qquad 
\delta^\alpha_\beta = P^{\alpha }_{\beta } + \sigma n^{\alpha }n_{\beta } \ .
\label{complete}
\end{equation}
For any vector $V \in \ct\left( \cm_5 \right) $, the vector $V^\alpha_\perp =
P^\alpha_\beta V^\beta $ is in $\ct\left( \Sigma \right), $
and so there is some vector $v \in \ct \left( \cm \right) $, such that

%
\begin{equation}
V_\perp^\alpha =v^\mu E_\mu^\alpha
\end{equation}

which entails

\begin{equation}
v_\mu = \gamma_{\mu\nu} v^\nu = g_{\alpha\beta} E^\alpha_\mu E^\beta_\nu v^\nu
= g_{\alpha\beta} E^\alpha_\mu V_\perp^\beta =  E^\alpha_\mu V^\perp_\alpha 
= E^\alpha_\mu P_\alpha^\beta V_\beta = E^\beta_\mu V_\beta
\end{equation}

since $ E^\alpha_\mu \in \ct\left( \Sigma \right)$.
In particular, expressing the metric in terms of (\ref{complete}), we find

\begin{equation}
\gamma _{\mu \nu }=g_{\alpha \beta
}E_{\mu }^{\alpha }E_{\nu }^{\beta }=\left( P_{\alpha \beta }+\sigma
n_{\alpha }n_{\beta }\right) E_{\mu }^{\alpha }E_{\nu }^{\beta }=P_{\alpha
\beta }E_{\mu }^{\alpha }E_{\nu }^{\beta } = P_{\mu \nu }
\end{equation}

so that the projector $P_{\alpha \beta }$ when restricted to $\Sigma $ acts
precisely as the 4D metric $\gamma_{\mu \nu }$.

Generalizing the characterization of 5D distance that is expressed in (\ref{difference}), 
we write 

\begin{equation}
X_2 - X_1 = \left( \delta x^\mu  + N^\mu \delta x^5 , N \delta x^5 \right)  
\qquad \qquad 
\delta X^\alpha = \left( \delta x^\mu  + N^\mu \delta x^5 \right) E^\alpha_\mu +
N n^\alpha \delta x^5
\label{difference-1}
\end{equation}

where $N$ is a lapse function and $N^\mu$ is a shift four-vector.
The 5D squared invariant interval now takes the~form
\begin{eqnarray}
dX^2 \eq  g_{\alpha\beta} \big( x,\tau\big)  \delta X^\alpha \delta X^\beta \notag \\
\eq  g_{\alpha\beta} \big( x,\tau\big)
\big[ \big( \delta x^\mu  + N^\mu \delta x^5 \big) E^\alpha_\mu +
 N n^\alpha \delta x^5\big] 
\big[ \big( \delta x^\nu  + N^\nu \delta x^5 \big) E^\beta_\nu +  N n^\beta \delta
x^5\big]  \notag \\ 
%
%
\eq \gamma_{\mu\nu} \big( x,\tau\big) \big( \delta x^\mu  + N^\mu \delta x^5 \big) 
\big( \delta x^\nu  + N^\nu \delta x^5 \big)
+ \sigma ^2 N^2 \big( \delta x^5\big) ^2 \notag \\
\eq \gamma_{\mu\nu} \big( x,\tau\big) \delta x^\mu \delta x^\nu 
+ 2 \gamma_{\mu\nu} \big( x,\tau\big) N^\nu \delta x^\mu \delta x^5 
+\textcolor{dRed}{\big( \gamma_{\mu\nu} \big( x,\tau\big) N^\mu N^\nu + \sigma ^2
N^2 \big) 
\big( \delta x^5\big) ^2 }
\label{5D-interval-2}
\end{eqnarray}
%


allowing us to decompose the 5D metric 
\vspace{-12pt}

\begin{equation}
g_{\alpha \beta } = \left[ 
\begin{array}{cc}
\gamma _{\mu \nu } & N_{\mu } \\ 
N_{\mu } & \sigma N^{2}+\gamma _{\mu \nu }N^{\mu }N^{\nu }
\end{array}
\right]
\qquad \qquad 
g^{\alpha \beta }=\left[ 
\begin{array}{cc}
\gamma ^{\mu \nu }+\sigma \dfrac{1}{N^{2}}N^{\mu }N^{\nu } & -\sigma \dfrac{1}{
N^{2}}N^{\mu } \strt{12} \\ 
-\sigma \dfrac{1}{N^{2}}N^{\mu } & \sigma \dfrac{1}{N^{2}}
\end{array}
\right]
\end{equation}

into the spacetime and $\tau$ sectors.
Once again, on any 4D \textcolor{dRed}{SHP spacetime $\cm (\tau)$}, the induced
metric $\gamma_{\mu\nu} ( x,\tau ) $ is just the local metric, we assumed to
exist at the outset. 
In this decomposition, the unit normal $n^\alpha$ becomes

\begin{equation}
n_\alpha = \sigma \frac{1}{\sqrt{\left\vert g^{55}\right\vert}} \partial_\alpha \p \tau(X)
=  \sigma N \delta^5_\alpha \ .
\label{unit_normal}
\end{equation}

One can easily establish that $\sqrt{g} = \sqrt{\gamma }N$ by writing 

\begin{equation}
\left[ 
\begin{array}{cc}
g_{\mu \nu } & g_{\mu 5} \\ 
g_{\mu 5} & g_{55}%
\end{array}%
\right] =\left[ 
\begin{array}{cc}
\gamma _{\mu \nu } & N_{\mu } \\ 
N_{\mu } & \sigma N^{2}+\gamma _{\mu \nu }N^{\mu }N^{\nu }%
\end{array}%
\right] 
= \left[ 
\begin{array}{cc}
I & 0 \\ 
N^{\mu } & 1%
\end{array}%
\right] \left[ 
\begin{array}{cc}
\gamma _{\mu \nu } & 0 \\ 
0 & \sigma N^{2}%
\end{array}%
\right] \left[ 
\begin{array}{cc}
I & N^{\nu } \\ 
0 & 1%
\end{array}%
\right] \ .
\label{E-330}
\end{equation}%

\subsection{Intrinsic and Extrinsic Geometry}

With compatible connection (\ref{connection}) the covariant derivative 
(\ref{cov-div}) on $\cm_5$ obeys $\nabla _\gamma  g_{\alpha \beta }  = 0$, leading~to the standard Ricci identity  

\begin{equation}
\left[ \nabla _{\beta } , \nabla _{\alpha } \right] X_{\delta } =
X_{\gamma }R_{\delta \alpha \beta }^{\gamma }
\label{F-curvature}
\end{equation}

with Riemann tensor

\begin{equation}
R_{\delta \alpha \beta }^{\gamma }=\frac{\partial }{\partial x^{\alpha }}
\Gamma _{\delta \beta }^{\gamma }-\frac{\partial }{\partial x^{\beta }}
\Gamma _{\delta \alpha }^{\gamma }+\Gamma _{\sigma \alpha }^{\gamma }\Gamma
_{\delta \beta }^{\sigma }-\Gamma _{\sigma \beta }^{\gamma }\Gamma _{\delta
\alpha }^{\sigma }  
\label{E-113}
\end{equation}

and associated Bianchi relations. 
To find the corresponding structures on the hyperspaces defined through foliation we
examine their projections onto $\ct(\Sigma)$.

For a vector $V=V^{\perp }\in \ct\left( \Sigma \right) $ we define the projected covariant
derivative $ \overline{\nabla }_{\alpha }$ in which the projected derivative acts on the
projected vector.  
Thus,

\begin{equation}
\overline{\nabla }_{\alpha }V_{\beta }^{\perp }=\overline{\nabla }_{\alpha
}\left( P_{\beta }^{\delta }V_{\delta }\right) =P_{\beta }^{\delta }
\overline{\nabla }_{\alpha }\left( V_{\delta }\right) =P_{\beta }^{\delta
}\left( P_{\alpha }^{\gamma }\nabla _{\gamma }\right) V_{\delta }=P_{\alpha
}^{\gamma }P_{\beta }^{\delta }\nabla _{\gamma }V_{\delta }  \ . 
\label{E-139}
\end{equation}

We justify the second equality by noting that the full 5D covariant derivative of
the projector is 

\begin{equation}
\nabla _{\alpha }P_{\beta \gamma }=\nabla _{\alpha }\left( g_{\beta \gamma
}-\sigma n_{\beta }n_{\gamma }\right) =-\sigma \nabla _{\alpha }\left(
n_{\beta }n_{\gamma }\right) =-\sigma \left[ \left( \nabla _{\alpha
}n_{\gamma }\right) n_{\beta }+n_{\beta }\nabla _{\alpha }n_{\gamma }\right]
\label{F-grad_P}
\end{equation}

and so the projected covariant derivative of the projector is 

\begin{eqnarray}
\overline{\nabla }_{\alpha }P_{\beta \gamma } \eq -\sigma P_{\alpha }^{\alpha
^{\prime }}P_{\beta }^{\beta ^{\prime }}P_{\gamma }^{\gamma ^{\prime
}}\left( \left( \nabla _{\alpha ^{\prime }}n_{\gamma ^{\prime }}\right)
n_{\beta ^{\prime }}+n_{\beta ^{\prime }}\nabla _{\alpha ^{\prime
}}n_{\gamma ^{\prime }}\right)  
\notag
\label{E-141} \\
\eq -\sigma P_{\alpha }^{\alpha ^{\prime }}P_{\gamma }^{\gamma ^{\prime
}}\left( \nabla _{\alpha ^{\prime }}n_{\gamma ^{\prime }}\right) \left(
P_{\beta }^{\beta ^{\prime }}n_{\beta ^{\prime }}\right) -\sigma P_{\alpha
}^{\alpha ^{\prime }}P_{\gamma }^{\gamma ^{\prime }}\left( P_{\beta }^{\beta
^{\prime }}n_{\beta ^{\prime }}\right) \nabla _{\alpha ^{\prime }}n_{\gamma
^{\prime }}=0  
\label{E-142}
\end{eqnarray}

which follows from $P_{\beta }^{\delta }n_{\delta }\equiv 0$.
This compatibility justifies regarding $\overline{\nabla }_{\alpha }$ as the intrinsic
covariant derivative on $\ct\left( \Sigma \right) $, denoted as

\begin{equation}
D_{\alpha }=\overline{\nabla }_{\alpha }=P_{\alpha }^{\gamma }\nabla
_{\gamma }\mbox{\qquad}D_{\mu }=E_{\mu }^{\alpha }D_{\alpha }=E_{\mu
}^{\alpha }P_{\alpha }^{\gamma }\nabla _{\gamma }=E_{\mu }^{\gamma }\nabla
_{\gamma }  
\label{E-143}
\end{equation}

and satisfying $ D_{\mu }\gamma _{\lambda \rho }=0$.
That is, for $V_\mu^\perp \in \ct\left( \Sigma \right) $ and $v_\nu \in \ct\left(
\cm\right) $ with $v_\mu = E_\mu^\alpha V_\alpha^\perp $
we have

\begin{equation}
E_{\mu }^{\alpha }E_{\nu }^{\beta }\left( D_{\alpha }V_{\beta }^{\perp
}\right) =E_{\mu }^{\alpha }D_{\alpha }E_{\nu }^{\beta }\left( P_{\beta
}^{\delta }V_{\delta }\right) =E_{\mu }^{\alpha }D_{\alpha }\left( E_{\nu
}^{\beta }P_{\beta }^{\delta }\right) V_{\delta }=E_{\mu }^{\alpha
}D_{\alpha }E_{\nu }^{\delta }V_{\delta }=D_{\mu }v_{\nu }  \ .
\label{E-148}
\end{equation}
%

%
The projected curvature $ \bar{R} _{\lambda \mu \nu }^{\rho }$ is defined through 

\begin{equation}
\left[ D_{\nu },D_{\mu }\right] X_{\lambda }=X_{\rho }\bar{R}
_{\lambda \mu \nu }^{\rho }  
\label{E-146}
\end{equation}

and will be examined below.

Restricted to $\ct\left( \Sigma \right) \subset \ct\left( \mathcal{M}
\right) $ 
the Weingarten map $\chi $ associates to a
tangent vector $V\in \ct\left( \Sigma \right) $ the variation of the $\tau$-like
unit vector $n$ along $V$. Thus,

\begin{equation}
\chi \left( V\right)
=\nabla _{V}n=V\cdot \left( \nabla n\right) 
\qquad \qquad 
\chi ^{\alpha }\left( V\right) = V^{\beta }\nabla _{\beta }n^{\alpha }
\end{equation}
%


and

\begin{equation}
U\cdot \left( \nabla _{V}n\right) =V\cdot \left( \nabla _{U}n\right) \ . 
\label{E-130}
\end{equation}

The extrinsic curvature on $\ct\left( \Sigma \right) $ is

\begin{equation}
K:\ct\left( \Sigma \right) \times \ct\left( \Sigma \right) \rightarrow R
\label{E-131}
\end{equation}

defined as the projection onto a vector $U$ of the Weingarten map along a vector $V$

\begin{eqnarray}
K\left( U,V\right) \eq -U\cdot \chi \left( V\right) =-U\cdot \nabla
_{V}n=-g_{\alpha \gamma }V^{\alpha }U^{\beta }\nabla _{\beta }n^{\gamma }
\label{E-132} \\
K_{\alpha \beta } \eq -g_{\alpha \gamma }\nabla _{\beta }n^{\gamma }=-\nabla
_{\beta }n_{\alpha }  \ . 
\label{E-133}
\end{eqnarray}

Using the projector $P_{\alpha \beta }$, we extend this definition to the full manifold
$\ct(\cm)$ as 

%
\begin{eqnarray}
K\left( U_{\perp },V_{\perp }\right) \eq K\left( PU,PV\right) =-g_{\gamma
\alpha }\left( P_{\varepsilon }^{\gamma }V^{\varepsilon }\right) \left(
P_{\phi }^{\beta }U^{\phi }\right) \nabla _{\beta }n^{\alpha }  
\label{E-135}
\\
K_{\phi \varepsilon }U^{\phi }V^{\varepsilon } \eq V^{\varepsilon }U^{\phi
}\left( -g_{\gamma \alpha }P_{\varepsilon }^{\gamma }P_{\phi }^{\beta
}\right) \nabla _{\beta }n^{\alpha }  
\label{E-136} \\
K_{\alpha \beta } \eq -P_{\alpha }^{\gamma }P_{\beta }^{\delta }~\nabla
_{\delta }n_{\gamma }  
\label{E-137}
\end{eqnarray}

where we recall that $\nabla _{\delta }n_{\gamma } $ may have both normal and tangent
components with respect to $\ct\left( \Sigma \right) $.
Because the projection is idempotent, we can write 
\begin{equation}
P_{\beta }^{\delta }\left( \nabla _{\gamma }n_{\delta }\right) \equiv \nabla
_{\gamma }n_{\beta }  
\label{E-170}
\end{equation}

leading to the identity

\begin{equation}
K_{\alpha \beta } =-P_{\alpha }^{\gamma }P_{\beta }^{\delta }\nabla
_{\gamma }n_{\delta }=-P_{\alpha }^{\gamma }\nabla _{\gamma }n_{\beta
}=-\left( \gamma _{\alpha }^{\gamma }-\sigma n_{\alpha }n^{\gamma }\right)
\nabla _{\gamma }n_{\beta }
= -\nabla _{\alpha }n_{\beta }+\sigma n_{\alpha }\left( n^{\gamma }\nabla
_{\gamma }n_{\beta }\right) 
\label{F-K_ident}
\end{equation}

and the contracted form

\begin{equation}
K=\gamma ^{\alpha \beta }K_{\alpha \beta }=\gamma ^{\alpha \beta }P_{\alpha
}^{\gamma }P_{\beta }^{\delta }\nabla _{\gamma }n_{\delta }=\gamma ^{\gamma
\delta }\nabla _{\gamma }n_{\delta }=\nabla _{\alpha }n^{\alpha } \ .
\label{E-155}
\end{equation}

Using (\ref{unit_normal}) for the unit normal $n_\alpha$, we may expand

\begin{eqnarray}
\left( n^{\gamma }\nabla _{\gamma }n_{\beta }\right)
 \eq \sigma n^{\gamma }\nabla _{\gamma }\left( N\nabla _{\beta
}\tau \right)
%
%
= \sigma n^{\gamma }\left( \nabla _{\gamma }N\right) \frac{n_{\beta }}{
\sigma N}+\sigma n^{\gamma }N\nabla _{\beta }\left( \frac{n_{\gamma }}{
\sigma N}\right)
\label{E-365} \notag \\
 \eq \frac{1}{N}\left[ n^{\gamma }n_{\beta }\nabla _{\gamma }N-\sigma \delta
_{\beta }^{\gamma }\nabla _{\gamma }N\right]
%
%
= -\sigma \frac{1}{N}\left[
\delta _{\beta }^{\gamma }-\sigma n^{\gamma }n_{\beta }\right] \nabla
_{\gamma }N
\label{E-368a} \notag \\
\eq -\sigma \frac{1}{N}P_{\beta }^{\gamma }\nabla _{\gamma }N
%
%
= -\sigma \frac{1}{N}D_{\beta }N 
\label{E-369}
\end{eqnarray}
%


to put (\ref{F-K_ident}) into the form

\begin{equation}
K_{\alpha \beta } = - \nabla_\alpha n_\beta - n_{\alpha }\frac{1}{N}D_{\beta }N
\ .
\label{F-K_ident-2}
\end{equation}

If $V \in \ct(\cm_5) $ has components both tangent and normal to
$\ct(\Sigma) $, and it so can be written as 

\begin{equation}
V^{\beta } = E_{\lambda }^{\beta }v^{\lambda }-\sigma \left( n\cdot
V\right) n^{\beta } \ \longrightarrow \
\nabla _{\alpha }V^{\textcolor{dRed}{\beta} } = \nabla _{\alpha }E_{\lambda }^{\beta
}v^{\lambda }-\sigma \nabla _{\alpha }\left( n\cdot V\right) n^{\beta }
\end{equation}

we see that

\begin{equation}
D_{\mu }v_{\nu }=E_{\mu }^{\alpha }E_{\nu }^{\beta }\nabla _{\alpha
}V_{\beta }-\sigma \left( n\cdot V\right) K_{\mu \nu }  
\label{E-162}
\end{equation}

in which the first term represents the tangential part of the covariant
derivative, and the second term is seen to expresses the connection for the normal
components of $V$ in the full covariant derivative. 

\subsection{Evolution of the Hypersurface $\Sigma $}

{From (\ref{difference-1}) we see that the variation of $X \in \Sigma$ for a
small time variation $\delta x^5$ at a given point $x_0 \in \cm$~is}   

\begin{equation}
\delta X^{\alpha }=\left( \dfrac{ \partial X^{\alpha }}{\partial
x^5}\right)_{x_0} \delta x^5 =\left( \dfrac{ \partial X^{\alpha }}{\partial
\tau}\right)_{x_0} \delta \tau \longrightarrow E^\alpha_5 = \left( \partial_5
\right)^\alpha = N n^\alpha + N^{\mu }E_{\mu }^{\alpha } 
\end{equation}

Defining $m^\alpha = N n^\alpha $ we write $E_5$ as $\partial_5 = m + {\mathbf N} $
and characterize time evolution through the Lie derivative in the time direction

\begin{equation}
{\mathcal{L}}_5 = {\mathcal{L}}_m + {\mathcal{L}}_{\mathbf N} \ .
\end{equation}

For the metric $ \gamma _{\alpha \beta }$, the Lie derivative is

\begin{equation}
{\mathcal{L}}_{m}\,\gamma _{\alpha \beta }  = m^{\gamma }\nabla _{\gamma
}\gamma _{\alpha \beta }+\gamma _{\gamma \beta }\nabla _{\alpha }m^{\gamma
}+\gamma _{\alpha \gamma }\nabla _{\beta }m^{\gamma }
\label{lie-1}
\end{equation}

which we may evaluate by using (\ref{projector}) for $ P_{\alpha \beta } =
\gamma_{\alpha \beta }$ in the first term and using (\ref{F-K_ident-2}) to obtain

\begin{equation}
\nabla _{\beta }m_{\alpha }= N\nabla _{\beta }n_{\alpha } +n_{\alpha } \nabla _{\beta } N = 
-NK_{\beta \alpha }-n_{\beta }D_{\alpha }N+n_{\alpha }\nabla _{\beta }Nu
\label{nab-m}
\end{equation}

in the remaining terms.  Notice that ${\mathcal{L}}_{m}\,P^\alpha_\beta$ is
the derivative in the normal direction of the projector onto the tangent space,
so that direct calculation while using (\ref{F-K_ident-2}) and (\ref{nab-m}) provides

\begin{equation}
{\mathcal{L}}_{m}\,P_{\ \,\beta }^{\alpha } 
= m^{\gamma }\nabla _{\gamma }(\delta^\alpha_\beta -\sigma n^\alpha
n_{\beta })-\gamma_{\ \,\beta }^{\gamma }\nabla _{\gamma }m^{\alpha
}+\gamma_{\ \,\gamma }^{\alpha }\nabla _{\beta }m^{\gamma } = 0
\end{equation}
%


expressing compatibility of ${\mathcal{L}}_{m}$ with $P_{\ \,\beta }^{\alpha }$. 
As a result, if $V \in \ct(\cm_5)$ is tangent to $\Sigma$, its Lie
derivative in the time direction is tangent to $\Sigma$, and so tangent vectors
propagate as tangent vectors as $\tau$ \textcolor{dRed}{advances monotonically}.  
As a result, (\ref{lie-1}) simplifies to

\begin{equation}
{\mathcal{L}}_{m}\,\gamma _{\alpha \beta } = -2NK_{\alpha \beta }
\end{equation}

leading to 

\begin{equation}
{\mathcal{L}}_5\,\gamma _{\alpha \beta } - 
{\mathcal{L}}_{{\mathbf N}}\,\gamma _{\alpha \beta } = -2NK_{\alpha \beta }
\ \longrightarrow \ 
{\mathcal{L}}_5\,\gamma _{\mu \nu } - {\mathcal{L}}_{{\mathbf N}}\,\gamma _{\mu \nu } = -2NK_{\mu \nu }
\end{equation}

as the evolution equation for the metric.

\subsection{Decomposition of the Riemann Tensor}
\label{de-R}

The 4+1 decomposition of $R_{\ \,\delta \alpha \beta }^{\gamma } $ is accomplished by
projecting onto $\Sigma $ and $n$.  Using the completeness relation (\ref{complete}) to
write 

\begin{equation}
R_{\ \,\delta \alpha \beta }^{\gamma }=\left( P_{\alpha }^{\alpha ^{\prime
}}+\sigma n_{\alpha }n^{\alpha ^{\prime }}\right) \left( P_{\beta }^{\beta
^{\prime }}+\sigma n_{\beta }n^{\beta ^{\prime }}\right) \left( P_{\gamma
^{\prime }}^{\gamma }{}\!+\sigma n^{\gamma }n_{\gamma ^{\prime }}\right)
\left( P_{\delta }^{\delta ^{\prime }}~+\sigma n_{\delta }n^{\delta ^{\prime
}}\right) R_{\ \,\delta ^{\prime }\alpha ^{\prime }\beta ^{\prime }}^{\gamma
^{\prime }}  
\label{E-397}
\end{equation}

we obtain products of the type

\begin{equation}
R_{\delta \alpha \beta }^{\gamma }= \delta_{\alpha }^{\alpha ^{\prime }}
\p\p \delta_{\beta }^{\beta ^{\prime }} \p\p  \delta_{\gamma ^{\prime }}^{\gamma }
\p\p \delta_{\delta }^{\delta ^{\prime }}\p\p  R_{\delta ^{\prime }\alpha ^{\prime }\beta
^{\prime }}^{\gamma ^{\prime }} \longrightarrow 
\left\{
\begin{array}{l}
E_\mu^\alpha 
\p\p E_\nu^\beta 
\p\p E_\gamma^\lambda
\p\p E_\sigma^\delta 
\p\p P_\alpha^{\alpha^\prime}
\p\p P_\beta^{\beta^\prime}
\p\p P_{\gamma^\prime}^\gamma
\p\p P_\delta^{\delta^\prime} \p\p
\p\p R_{\delta^\prime \alpha^\prime \beta^\prime \gamma^\prime}
= R_{\sigma \mu \nu }^\lambda \strt{12} \\
E_\mu^\alpha
\p\p E_\nu^\beta
\p\p E_\gamma^\lambda
\p\p P_{\gamma^\prime}^\gamma n^\delta
\p\p P_\alpha^{\alpha^\prime}
\p\p P_\beta^{\beta^\prime}
\p\p R_{\delta \alpha^\prime \beta^\prime}^{\gamma^\prime} 
= \sigma N \p\p R_{5\mu \nu }^\lambda \strt{12} \\
E^{\alpha \mu }
\p\p E_\nu^\beta
\p\p P_{\alpha \alpha^\prime} \p\p n^\delta
\p\p P_\beta^{\beta^\prime} \p\p n^\gamma
\p\p R_{\delta \beta^\prime \gamma}^{\alpha^\prime} = N^2 \p\p R_{5\nu 5}^\mu
\end{array}
\right.
\end{equation}

where $ \ R_{\delta \alpha \beta }^{\gamma } \p\p n^\delta \p\p n^\alpha \p\p
n^\beta = 0 \ $, because of the symmetries of the Riemann tensor.
For the projected curvature defined in (\ref{E-146}), we write

\begin{equation}
D_{\alpha }D_{\beta }V^{\gamma }=D_{\alpha }\left( D_{\beta }V^{\gamma
}\right) =P_{\alpha }^{\alpha ^{\prime }}P_{\beta }^{\beta ^{\prime
}}P_{\gamma }^{\gamma ^{\prime }}\nabla _{\alpha ^{\prime }}\left( D_{\beta
^{\prime }}V^{\gamma ^{\prime }}\right)  
\label{E-183}
\end{equation}

we expand and use (\ref{F-grad_P}) in order to obtain

\begin{equation}
D_{\alpha }D_{\beta }V^{\gamma }=\sigma K_{\alpha \beta }P_{\gamma ^{\prime
}}^{\gamma }n^{\beta ^{\prime }}\nabla _{\beta ^{\prime }}V^{\gamma ^{\prime
}}+\sigma K_{\alpha }^{\gamma }K_{\beta \delta }V^{\delta }+P_{\alpha
}^{\alpha ^{\prime }}P_{\beta }^{\beta ^{\prime \prime }}P_{\gamma ^{\prime
\prime }}^{\gamma }(\nabla _{\alpha ^{\prime }}\nabla _{\beta ^{\prime
\prime }}V^{\gamma ^{\prime \prime }})  
\label{E-206}
\end{equation}

so that

\begin{equation}
\left[ D_{\alpha },D_{\beta }\right] V^{\gamma }=
\bar{R}_{\delta \alpha \beta }^{\gamma }V^{\delta } =
-\sigma \left(
K_{\alpha \delta }K_{\ \,\beta }^{\gamma }-K_{\beta \delta }K_{\ \,\alpha
}^{\gamma }\right) V^{\delta }+P_{\ \,\alpha }^{\alpha ^{\prime }}P_{\
\,\beta }^{\beta ^{\prime }}P_{\ \,\gamma ^{\prime }}^{\gamma }{}\!R_{\
\,\delta ^{\prime }\alpha ^{\prime }\beta ^{\prime }}^{\gamma ^{\prime
}}P_{\ \,\delta }^{\delta ^{\prime }}V^{\delta }
\label{E-208}
\end{equation}

which, by the quotient theorem on $\Sigma $, leads to 

\begin{equation}
P_{\ \,\alpha }^{\alpha ^{\prime }}P_{\ \,\beta }^{\beta ^{\prime }}P_{\
\,\gamma ^{\prime }}^{\gamma }{}\!P_{\ \,\delta }^{\delta ^{\prime }}R_{\
\,\delta ^{\prime }\alpha ^{\prime }\beta ^{\prime }}^{\gamma ^{\prime }} =
\bar{R}_{\delta \alpha \beta }^{\gamma }-\sigma \left( K_{\alpha }^{\gamma
}K_{\beta \delta }-K_{\beta }^{\gamma }K_{\alpha \delta }\right)
\label{F-gauss}
\end{equation}

This is known as the Gauss relation. 
Acting on this expression with
$E_{\gamma }^{\mu }E_{\nu }^{\delta }E_{\lambda }^{\alpha }E_{\rho }^{\beta
} $
we find

\begin{equation}
R_{\ \,\nu \lambda \rho }^{\mu }=\bar{R}_{\ \,\nu \lambda \rho }^{\mu
}-\sigma \left( K_{\lambda }^{\mu }K_{\rho \nu }-K_{\rho }^{\mu }K_{\lambda
\nu }\right)  
\label{E-216}
\end{equation}

providing an expression for the intrinsic curvature
$R_{\ \,\nu \lambda \rho }^{\mu } $ in terms of the projected curvature $ \bar{R}_{\
\,\nu \lambda \rho }^{\mu }$ and the intrinsic curvature $ K_{\rho \nu }$. 
Contracting on $\alpha $ and $\gamma $ in (\ref{F-gauss}) leads to

\begin{equation}
P_{\ \,\alpha }^{\alpha ^{\prime }}P_{\ \,\beta }^{\beta ^{\prime }}R_{\
\,\alpha ^{\prime }\beta ^{\prime }}-\sigma P_{\alpha \alpha ^{\prime
}}n^{\delta ^{\prime }}P_{\ \,\beta }^{\beta ^{\prime }}n^{\gamma \prime
}R_{\ \,\delta ^{\prime }\beta ^{\prime }\gamma ^{\prime }}^{\alpha ^{\prime
}}=\bar{R}_{\alpha \beta }-\sigma \left( KK_{\alpha \beta }-K_{\alpha
}^{\delta }K_{\beta \delta }\right)  
\label{E-221}
\end{equation}

and contracting on $\alpha $ and $\beta $ gives

\begin{equation}
R-2\sigma R_{\alpha \beta }n^{\alpha }n^{\beta } = \bar{R}-\sigma
\left( K^{2}-K^{\alpha \beta }K_{\alpha \beta }\right)
\label{F-scalar_gauss}
\end{equation}

called the scalar Gauss relation.

Applying the Ricci identity (\ref{F-curvature}) to the vector $n$ as

\begin{equation}
\left( \nabla _{\beta }\nabla _{\alpha }-\nabla _{\alpha }\nabla _{\beta
}\right) n^{\gamma }=R_{\gamma ^{\prime }\alpha \beta }^{\gamma }n^{\gamma
^{\prime }}  
\label{E-228}
\end{equation}

projecting the LHS onto $\Sigma $ as

\begin{equation}
P_{\ \,\alpha }^{\alpha ^{\prime }}P_{\ \,\beta }^{\beta ^{\prime
}}P_{\gamma ^{\prime }}^{\gamma }\left( \nabla _{\alpha ^{\prime }}\nabla
_{\beta ^{\prime }}-\nabla _{\beta ^{\prime }}\nabla _{\alpha ^{\prime
}}\right) n^{\gamma ^{\prime }}  
\label{E-229}
\end{equation}

and using the identity (\ref{F-K_ident}) leads us to 

\begin{equation}
D_{\beta }K_{\alpha }^{\gamma }-D_{\alpha }K_{\beta }^{\gamma }=P_{\gamma
^{\prime }}^{\gamma }n^{\delta }P_{\ \,\alpha }^{\alpha ^{\prime }}P_{\
\,\beta }^{\beta ^{\prime }}R_{\delta \alpha ^{\prime }\beta ^{\prime
}}^{\gamma ^{\prime }}  
\label{F-Codazzi}
\end{equation}

which is called the Codazzi relation.  
Contracting on $ \alpha$ and $ \gamma$ produces

\begin{equation}
n_{\delta }R_{\mu \nu \lambda }^{\delta }=D_{\lambda }K_{\nu \mu }-D_{\nu
}K_{\lambda \mu }  \ .
\label{E-239}
\end{equation}

Using (\ref{unit_normal}) for the unit normal $n_\alpha$ provides an
interpretation of this expression as

\begin{equation}
n_{\delta }R_{\mu \nu \lambda }^{\delta }=\sigma N\delta _{\delta
}^{5}R_{\mu \nu \lambda }^{\delta }\longrightarrow R_{\mu \nu \lambda
}^{5}=\sigma \frac{1}{N}\left( D_{\lambda }K_{\nu \mu }-D_{\nu }K_{\lambda
\mu }\right)  
\label{E-252}
\end{equation}

recalling the role of the extrinsic curvature $ K_{\mu \nu }$ as the curvature of $\cm $
mapped to the hypersurface $\Sigma $ and embedded in the larger manifold $\cm_5 $.

Returning to the Ricci identity for $n^\alpha$, we apply (\ref{F-K_ident-2}) twice
to terms $\nabla _{\beta }\nabla _{\gamma }n^{\alpha }$
and project onto~(\ref{E-228}) with
$P_{\alpha \alpha ^{\prime }}n^{\gamma ^{\prime }}P_{\beta }^{\beta ^{\prime }}$ to obtain

\begin{equation}
-K_{\alpha \gamma }K_{\ \,\beta }^{\gamma }+\frac{1}{N}D_{\beta }D_{\alpha
}N + 
P_{\ \,\alpha }^{\alpha^\prime }P_{\ \,\beta }^{\beta^\prime }\,n^{\gamma }\nabla
_{\gamma }K_{\alpha^\prime \beta^\prime }  
=
P_{\alpha \alpha ^{\prime }}n^{\gamma ^{\prime }}P_{\beta }^{\beta ^{\prime
}}~R_{\delta \beta ^{\prime }\gamma ^{\prime }}^{\alpha ^{\prime }}n^{\delta
} \ .
\label{E-406}
\end{equation}

Again using (\ref{nab-m}) in the Lie derivative of $K_{\alpha \beta }$
to write

\begin{equation}
{\mathcal{L}}_{m}\,K_{\alpha \beta }=Nn^{\gamma }\nabla _{\gamma }K_{\alpha
\beta }-2NK_{\alpha \gamma }K_{\ \,\beta }^{\gamma }-K_{\alpha \gamma
}D^{\gamma }Nn_{\beta }-K_{\beta \gamma }D^{\gamma }Nn_{\alpha }
\label{E-409}
\end{equation}

the last two equations are combined as

\begin{equation}
\frac{1}{N}{\mathcal{L}}_{m}\,K_{\alpha \beta }+\frac{1}{N}D_{\alpha }D_{\beta
}N+K_{\alpha \gamma }K_{\ \,\beta }^{\gamma }  = P_{\alpha \alpha ^{\prime
}}\,n^{\delta }P_{\ \,\beta }^{\beta ^{\prime }}\,n^{\gamma }\,\!R_{\ \,\delta
\beta ^{\prime }\gamma }^{\alpha ^{\prime }} 
\label{F-proj3}
\end{equation}

to provide an evolution equation for $K_{\alpha \beta }$.  
Rewriting (\ref{E-221}) as

\begin{equation}
P_{\alpha \alpha ^{\prime }}n^{\delta ^{\prime }}P_{\ \,\beta }^{\beta
^{\prime }}n^{\gamma \prime }R_{\ \,\delta ^{\prime }\beta ^{\prime }\gamma
^{\prime }}^{\alpha ^{\prime }} = \sigma P_{\ \,\alpha }^{\alpha ^{\prime
}}P_{\ \,\beta }^{\beta ^{\prime }}R_{\ \,\alpha ^{\prime }\beta ^{\prime }}
-\sigma \bar{R}_{\alpha \beta }+KK_{\alpha \beta }-K_{\alpha }^{\delta
}K_{\beta \delta }
\end{equation}

we can put (\ref{F-proj3}) into the form

\begin{equation}
P_{\ \,\alpha }^{\alpha ^{\prime }}P_{\ \,\beta }^{\beta ^{\prime }}R_{\
\,\alpha ^{\prime }\beta ^{\prime }} = \sigma
\frac{1}{N}{\mathcal{L}}_{m}\,K_{\alpha \beta }+\sigma \frac{1}{N} D_{\alpha
}D_{\beta }N+\bar{R}_{\alpha \beta }-\sigma KK_{\alpha \beta }+\sigma 2K_{\alpha
}^{\delta }K_{\beta \delta }   
\label{F-ricci_proj}
\end{equation}

in which only $P_{\ \,\alpha }^{\alpha ^{\prime }}P_{\ \,\beta }^{\beta ^{\prime }}R_{\
\,\alpha ^{\prime }\beta ^{\prime }}$ on the LHS refers to the 5D geometry of
$\cm_5$.

\subsection{Decomposition of the Einstein Equation}

The Einstein equations 

\begin{equation}
G_{\alpha \beta }=R_{\alpha \beta }-\frac{1}{2}g_{\alpha \beta }R=\frac{8\pi
G}{c^{4}}T_{\alpha \beta }  
\label{E-446}
\end{equation}

can be written

\begin{equation}
R_{\alpha \beta }=\frac{8\pi G}{c^{4}}\left( T_{\alpha \beta }-\frac{1}{2}
g_{\alpha \beta }T\right)  
\label{E-447}
\end{equation}

where $T=g^{\alpha \beta }T_{\alpha \beta }$. 
As above, we decompose the field equations by projecting onto $\Sigma $ and $n$
as

\begin{equation}
T_{\alpha \beta } = T_{\alpha ^{\prime }\beta ^{\prime }}
\left( P_{\alpha }^{\alpha ^{\prime }}+\sigma n^{\alpha ^{\prime }}n_{\alpha }\right) \left(
P_{\beta }^{\beta ^{\prime }}+\sigma n^{\beta ^{\prime }}n_{\beta }\right)
= S_{\alpha \beta }+2\sigma n_{\alpha }p_{\beta }+n_{\alpha }n_{\beta }\kappa
\end{equation}

where 

\begin{equation}
S_{\alpha \beta }  = P_{\alpha }^{\alpha ^{\prime }}P_{\beta }^{\beta
^{\prime }}T_{\alpha ^{\prime }\beta ^{\prime }}
\qquad \qquad 
p_{\beta }  = -n^{\alpha ^{\prime }}P_{\beta }^{\beta ^{\prime }}T_{\alpha
^{\prime }\beta ^{\prime }}
\qquad \qquad 
\kappa  = n^{\alpha }n^{\beta }T_{\alpha \beta }
\end{equation}
%

so that $ S_{\mu \nu }$ corresponds to \textcolor{dRed}{the 4D
energy-momentum tensor} $ T_{\mu \nu }$, $ p_{\mu }$ corresponds
to \textcolor{dRed}{the mass current into the $\mu$ direction} $ T_{5\mu }$,
and $ \kappa$ corresponds to \textcolor{dRed}{the mass density} $ T_{55}$.
\textcolor{dRed}{It is useful to regard mass in this context as being related to the
difference between energy and momentum, a dynamical quantity in the SHP
framework.} 
The trace is

\begin{equation}
T  = g^{\alpha \beta }T_{\alpha \beta }=g^{\alpha \beta }\left( S_{\alpha
\beta }-2\sigma n_{\alpha }p_{\beta }+n_{\alpha }n_{\beta }\kappa \right)
 = S-2\sigma g^{\alpha \beta }n_{\alpha }p_{\beta }+g^{\alpha \beta
}n_{\alpha }n_{\beta }\kappa  
 = S+\sigma \kappa  
\label{E-458}
\end{equation}

where we used

\begin{equation}
g^{\alpha \beta }n_{\alpha }p_{\beta }=n\cdot p=0  
\label{E-459}
\end{equation}

which follows from 

\begin{equation}
p_{\beta }= -P_{\beta }^{\beta ^{\prime }}\left( n^{\alpha }T_{\alpha \beta
^{\prime }}\right) \in \ct\left( \Sigma \right)  \ . 
\label{E-460}
\end{equation}

Thus, projecting the field equations (\ref{E-447}) onto $\Sigma$ 
with $P_{\alpha }^{\alpha ^{\prime
}}P_{\beta }^{\beta ^{\prime }} $ leads to

\begin{equation}
P_{\alpha }^{\alpha ^{\prime }}P_{\beta }^{\beta ^{\prime }}\left( T_{\alpha
^{\prime }\beta ^{\prime }}-\frac{1}{2}g_{\alpha ^{\prime }\beta ^{\prime
}}T\right) =S_{\alpha \beta }-\frac{1}{2}\gamma _{\alpha \beta }\left(
S+\sigma \kappa \right)
\end{equation}

on the RHS and while the LHS is $P_{\ \,\alpha }^{\alpha ^{\prime }}P_{\
\,\beta }^{\beta ^{\prime }}R_{\ \,\alpha ^{\prime }\beta ^{\prime }}$ which
from (\ref{F-ricci_proj}) provides

\begin{equation}
{\mathcal{L}}_{m}\,K_{\mu \nu }=-D_{\mu }D_{\nu }N+N\left\{ -\sigma \bar{R}
_{\mu \nu }+KK_{\mu \nu }-2K_{\mu }^{\lambda }K_{\nu \lambda }+\sigma \frac{
8\pi G}{c^{4}}\left[ S_{\mu \nu }-\frac{1}{2}\gamma _{\mu \nu }\left(
S+\sigma \kappa \right) \right] \right\}  
\label{F-evolve}
\end{equation}

as the evolution equation for $K_{\mu \nu }$.

The double projection onto the time direction $n$ is

\begin{equation}
\left( R_{\alpha \beta }-\frac{1}{2}g_{\alpha \beta }R\right) n^{\alpha
}n^{\beta }  = \frac{8\pi G}{c^{4}}T_{\alpha \beta }n^{\alpha }n^{\beta }
\qquad \longrightarrow \qquad 
R_{\alpha \beta }n^{\alpha }n^{\beta }-\frac{1}{2}\sigma R  = \frac{8\pi G}{
c^{4}}\kappa
\label{E-465} 
\end{equation}
%

and using the scalar Gauss relation (\ref{F-scalar_gauss}), we obtain

\begin{equation}
\bar{R}-\sigma \left( K^{2}-K^{\mu \nu }K_{\mu \nu }\right)  = -\sigma \frac{
16\pi G}{c^{4}}\kappa  
\label{E-471}
\end{equation}

This expression, called the Hamiltonian constraint, has no $\tau $-derivatives,
and so, if it is satisfied by the initial conditions, then it will be satisfied at all times.
\textcolor{dRed}{We observe that this constraint applies to the mass density of
the gravitational field, not the energy density as in 4D GR.}



The mixed projection with $P_{\ \,\beta }^{\beta ^{\prime }}n^{\alpha } $

\begin{equation}
n^{\alpha }P_{\ \,\beta }^{\beta ^{\prime }}\left( R_{\alpha \beta ^{\prime
}}-\frac{1}{2}g_{\alpha \beta ^{\prime }}R\right) =n^{\alpha }P_{\ \,\beta
}^{\beta ^{\prime }}\frac{8\pi G}{c^{4}}T_{\alpha \beta ^{\prime }}
\longrightarrow
P_{\ \,\beta }^{\beta ^{\prime }}n^{\alpha }R_{\alpha \beta ^{\prime }}-
\frac{1}{2}g_{\alpha \beta ^{\prime }}n^{\alpha }P_{\ \,\beta }^{\beta
^{\prime }}R  = -\frac{8\pi G}{c^{4}}p_{\beta }  
\label{E-475}
\end{equation}

is combined with the contracted Codazzi relation (\ref{E-239}) and 
$g_{\alpha \beta ^{\prime }}n^{\alpha }P_{\ \,\beta }^{\beta ^{\prime
}}=n^{\alpha }P_{\alpha \beta }=0$ to obtain

\begin{equation}
D_{\mu }K_{\nu }^{\mu }-D_{\nu }K  = \frac{8\pi G}{c^{4}}p_{\nu }
\label{E-476}
\end{equation}
%


which is called the momentum constraint, \textcolor{dRed}{referring to the flow
of mass into the field}, and it also has no $\tau $-derivatives.
We notice that the evolution equation contains only objects defined on $\Sigma $
and, thus, includes no factors of $\Gamma _{\mu \nu }^{5}$. 
Any such factors can only appear in the constraint equations. 

Writing the Einstein tensor as 

\begin{equation}
G_{\textcolor{dRed}{\alpha \beta} } = R_{\alpha \beta }-\frac{1}{2}g_{\alpha \beta }R
\end{equation}

the Bianchi relations are 

\begin{equation}
\nabla _{\alpha }G^{\alpha \beta }=\partial _{\alpha }G^{\alpha \beta }+
\text{Christoffel Symbols}\times G^{\alpha \beta }=0  
\label{E-477}
\end{equation}

forming a set of relations among the field entities.
The five equations in 5D reduce the number of independent components of
$G^{\alpha \beta }$ from fifteen to ten, and are understood as constraints on the
initial conditions of the evolution equations. 
Because the Einstein equations are second order in $\tau$ derivatives of the
metric, a solution requires that the initial conditions include $\gamma_{\mu \nu
}$ and $\partial_{\tau }\gamma_{\mu \nu }$ at the initial time. 
Expanding $\partial _{\alpha }G^{\alpha \beta }  = \partial_{\mu }G^{\mu
\beta } + \partial_{5 }G^{5 \beta }$ to rewrite the Bianchi relations as 

\begin{equation}
\frac{1}{c_5}\partial _{\tau }G^{5\beta }=-\partial _{\mu }G^{\mu \beta }-\text{
Christoffel Symbols}\times G^{\alpha \beta }  
\label{E-478}
\end{equation}

and noticing that the LHS cannot be more than second order in $\partial
_{\tau }$, we see that $G^{5\beta }$ cannot be more than first
order in $\partial _{\tau }$.  
Therefore, the expressions contained in $G^{5\beta }$ must be part of the initial conditions.
The Einstein equations

\begin{equation}
G^{\alpha \beta }=8\pi G_{N}T^{\alpha \beta }  
\label{E-479}
\end{equation}

thus split into components

\begin{eqnarray}
G^{\mu \beta }  \eq 0 \ \longrightarrow \ \text{ten equations of second order in
} \partial_\tau
\label{E-480} \\
G^{5\beta }  \eq 0 \ \longrightarrow \ \text{five relations among the
initial conditions of first order } \partial_\tau  \ .
\label{E-481}
\end{eqnarray}

Moreover, the constraints are propagated to future times because

\begin{equation}
G^{5\beta }|_{\tau _{0}}=0\Rightarrow \partial _{\beta }G^{5\beta }|_{\tau
_{0}}=0\Rightarrow \partial _{\tau }G^{5\beta }|_{\tau _{0}}=0  
\label{E-482}
\end{equation}

and so they do not change.


\subsection{Summary of Einstein System as Differential Equations}

The decomposition of the Einstein equations into a 4+1 system of partial
differential equations permits particular structures to be solved as an initial
value problem.
The initial conditions that are to be specified at some $\tau$ are the metric
$\gamma_{\mu \nu }$ and its first Lie derivative $ K_{\mu \nu }$, along with the
mass-energy-momentum distribution of matter as represented by $S_{\mu \nu } $,
$p_{\nu }$, and $\kappa $.
The initial conditions must satisfy the Hamiltonian constraint,
\textcolor{dRed}{a constraint on mass rather than energy,}

\begin{equation}
\bar{R}-\sigma \left( K^{2}-K^{\mu \nu }K_{\mu \nu }\right)  = -\sigma \frac{
16\pi G}{c^{4}}\kappa
\label{einstein-3}
\end{equation}

and the momentum constraint

\begin{equation}
D_{\mu }K_{\nu }^{\mu }-D_{\nu }K  = \frac{8\pi G}{c^{4}}p_{\nu }
\label{einstein-4}
\end{equation}

at the initial time.
Because these constraints contain no $\tau$ derivatives, they are guaranteed to
be satisfied at subsequent times.
Given appropriate initial conditions, the metric is found at subsequent times by
integrating forward---analytically or numerically---the coupled evolution equations

\begin{equation}
\frac{1}{c_5}{\mathcal{L}}_\tau\,\gamma _{\mu \nu } -
{\mathcal{L}}_{{\mathbf N}}\,\gamma _{\mu \nu } = -2NK_{\mu \nu }
\label{einstein-1}
\end{equation}
and

\begin{eqnarray}
\left( \frac{1}{c_5}{\mathcal{L}}_\tau - {\mathcal{L}}_{{\mathbf N}} \right)K_{\mu \nu } 
\eq -D_{\mu }D_{\nu }N \vspace{-4pt} \notag  \\ && \hspace{-44pt} +N\left\{ -\sigma \bar{R}
_{\mu \nu }+KK_{\mu \nu }-2K_{\mu }^{\lambda }K_{\nu \lambda }+\sigma \frac{
8\pi G}{c^{4}}\left[ S_{\mu \nu }-\frac{1}{2}\gamma _{\mu \nu }\left(
S+\sigma \kappa \right) \right] \right\}
\label{einstein-2}
\end{eqnarray}

We note that the lapse $N$ and shift $N^\mu$ are not dynamical variables,
but they are part of the metric specified at the initial time.

\section{The ADM Hamiltonian Formulation}
\label{ADM}


\textcolor{dRed}{The configuration space variable in the ADM
formalism is 
$g _{\alpha \beta }=g _{\alpha \beta }\left( \gamma _{\mu \nu
},N^{\mu },N\right)$,
the full metric on $\cm_5$.
Becausre $N$ and $N^\mu$ are not dynamical, the phase space consists of $\gamma _{\mu \nu }$ and 
$\dot \gamma _{\mu \nu }$, as given by}

\begin{equation}
\dot \gamma _{\mu \nu } = \frac{1}{c_5}{\mathcal{L}}_\tau\,\gamma _{\mu \nu }  = 
{\mathcal{L}}_{{\mathbf N}}\,\gamma _{\mu \nu }+2NK_{\mu \nu }
\label{einstein-1a}
\end{equation}
where we use (\ref{einstein-1}) with the sign convention for $K_{\mu \nu }$ reversed.
Contracting on $\alpha $ and $\beta $ in (\ref{F-ricci_proj}) and combining with
the scalar Gauss relation (\ref{F-scalar_gauss}), we obtain 

\begin{equation}
R=\bar{R}-\sigma \left( K^{2}-K^{\alpha \beta }K_{\alpha \beta }\right)
+2\sigma \nabla _{\alpha }\left( n^{\beta }\nabla _{\beta }n^{\alpha
}-n^{\alpha }\nabla _{\beta }n^{\beta }\right)  
\label{E-519}
\end{equation}

so that the Einstein-Hilbert action for GR in the absence of matter can be expanded as

\begin{equation}
S_{\textcolor{dRed}{ADM}}\left[ \gamma _{\mu \nu },\dot{\gamma}_{\mu \nu },N^{\mu },N\right]
=\int d\tau d^{4}x~\sqrt{g} \bar{R}
=\int d\tau d^{4}x~\sqrt{\gamma }N\left[ \bar{R}-\sigma \left( K^{\mu \nu }K_{\mu
\nu }-K^{2}\right) \right]
\label{E-523}
\end{equation}

where $g = \left\vert \det g_{\alpha \beta} \right\vert$ and
$\gamma = \left\vert \det \gamma_{\mu \nu } \right\vert$
and the total gradient in (\ref{E-519}) is discarded as a boundary term.
The DeWitt metric is defined as

\begin{equation}
G^{\mu \nu \lambda \rho }=\frac{1}{2}\left( \gamma ^{\mu \lambda }\gamma
^{\nu \rho }+\gamma ^{\mu \rho }\gamma ^{\nu \lambda }-2\gamma ^{\mu \nu
}\gamma ^{\lambda \rho }\right)  
\label{E-524}
\end{equation}

with inverse in $D$ dimensions

\begin{equation}
G_{\mu \nu \lambda \rho } = \frac{1}{2}\left( \gamma
_{\lambda \zeta }\gamma _{\rho \kappa }+\gamma _{\lambda \kappa }\gamma
_{\rho \zeta }-\frac{2}{D-1}\gamma _{\lambda \rho }\gamma _{\zeta \kappa
}\right)
\end{equation}

in terms of which

\begin{equation}
G^{\mu \nu \lambda \rho }K_{\mu \nu }K_{\lambda \rho } = \frac{1}{2}\left(
\gamma ^{\mu \lambda }\gamma ^{\nu \rho }+\gamma ^{\mu \rho }\gamma ^{\nu
\lambda }-2\gamma ^{\mu \nu }\gamma ^{\lambda \rho }\right) K_{\mu \nu
}K_{\lambda \rho }  
= K^{\mu \nu }K_{\mu \nu }-K^{2}  
\label{E-527}
\end{equation}

so that 

\begin{equation}
\mathcal{L}_{\textcolor{dRed}{ADM}}\left[ \gamma _{\mu \nu },\dot{\gamma}_{\mu \nu },N^{\mu },N
\right] =\sqrt{\gamma }N\left[ -\sigma G^{\mu \nu \lambda \rho }K_{\mu \nu
}K_{\lambda \rho }+\bar{R}\right]  \ .
\label{E-528}
\end{equation}

Because $K^{\mu \nu }$ is first order in derivatives, the first term has
the form of kinetic energy. The canonical conjugate momentum to 
$\gamma _{\mu \nu }$ is 

\begin{equation}
\pi ^{\mu \nu }=\frac{\partial \mathcal{L}_{\textcolor{dRed}{ADM}}}{\partial \dot{\gamma}_{\mu
\nu }}=-2\sigma \sqrt{\gamma }NG^{\zeta \kappa \lambda \rho }K_{\lambda \rho
}\frac{\partial K_{\zeta \kappa }}{\partial \dot{\gamma}_{\mu \nu }}
\label{E-529}
\end{equation}

so that using (\ref{einstein-1a}) to obtain

\begin{equation}
\frac{\partial
K_{\zeta \kappa }}{\partial \dot{\gamma}_{\mu \nu }}=\frac{1}{2N}\delta
_{\zeta }^{\mu }\delta _{\kappa }^{\nu }  
\label{E-531}
\end{equation}

we find

\begin{equation}
\pi ^{\mu \nu }=-2\sigma \sqrt{\gamma }NG^{\zeta \kappa \lambda \rho
}K_{\lambda \rho }\frac{1}{2N}\delta _{\zeta }^{\mu }\delta _{\kappa }^{\nu
}=-\sigma \sqrt{\gamma }G^{\mu \nu \lambda \rho }K_{\lambda \rho }=-\sigma 
\sqrt{\gamma }\left( K^{\mu \nu }-\gamma ^{\mu \nu }K\right)  
\label{E-532}
\end{equation}

with trace 

\begin{equation}
\pi  = \gamma _{\mu \nu }\pi ^{\mu \nu }=\sigma \left( D-1\right) \sqrt{ \gamma }K  
\ \ \longrightarrow \ \
K  = \frac{\sigma }{\left( D-1\right) \sqrt{\gamma }}\pi  \ .
\label{E-534}
\end{equation}

Writing $K^{\mu \nu }$ in terms of $\pi ^{\mu \nu }$ 

\begin{equation}
K^{\mu \nu }  = -\frac{\sigma }{\sqrt{\gamma }}\left( \pi ^{\mu \nu }-\gamma
^{\mu \nu }\frac{1}{\left( D-1\right) }\pi \right)  
\label{E-538}
\end{equation}

and lowering the indices of $\pi^{\mu \nu }$

\begin{equation}
G_{\mu \nu \lambda \rho }\pi ^{\lambda \rho } = 
\pi _{\mu \nu }-\frac{1}{D-1}\gamma _{\mu \nu }\pi
= -\sigma \sqrt{\gamma }K_{\mu \nu }
\label{E-558}
\end{equation}

we see that $K_{\mu \nu }$ represents the momentum conjugate to $
\gamma _{\mu \nu }$.
Replacing $K_{\mu \nu }$ in (\ref{einstein-1a}), we can write the velocity as

\begin{equation}
\dot{\gamma}_{\mu \nu }  = 
-\sigma \frac{2N}{\sqrt{\gamma }}G_{\mu \nu \lambda \rho }\pi ^{\lambda
\rho }+L_{N}\gamma _{\mu \nu }  
\label{E-564}
\end{equation}

in terms of the momentum and configuration variable.

Because $\bar R$ is independent of the lapse $N$ and shift $N^{\mu }$ the
Lagrangian $\mathcal{L}_{\textcolor{dRed}{ADM}}\left[ \gamma _{\mu \nu },\dot{\gamma}_{\mu \nu
},N^{\mu },N\right] $ contains no derivatives of $N,N^{\mu }$ and these
act as Lagrange multipliers enforcing as constraints their conjugates.
Thus

\begin{equation}
p_{N}=\frac{\partial \mathcal{L}_{\textcolor{dRed}{ADM}}}{\partial \dot{N}}\mbox{\qquad}
p_{N^{\mu }}=\frac{\partial \mathcal{L}_{\textcolor{dRed}{ADM}}}{\partial \dot{N}^{\mu }}=0
\label{E-565}
\end{equation}

and variation with respect to the lapse and shift produces

\begin{equation}
0 
=-\frac{\partial }{
\partial N}\left( \sqrt{\gamma }N\left[ R-\sigma G^{\mu \nu \lambda \rho
}K_{\mu \nu }K_{\lambda \rho }\right] \right)  
 = -\sqrt{\gamma }R+\sigma \sqrt{\gamma }G^{\mu \nu \lambda \rho }\frac{
\partial }{\partial N}\left( NK_{\mu \nu }K_{\lambda \rho }\right) \ .
\label{E-567}
\end{equation}

Rewriting (\ref{einstein-1a}) as 

\begin{equation}
K_{\mu \nu }=\frac{1}{2N}\left( \dot{\gamma}_{\mu \nu }-D_{\mu }N_{\nu
}+D_{\nu }N_{\mu }\right)  
\ \ \longrightarrow \ \ NK_{\mu \nu }K_{\lambda \rho }\sim \frac{1}{N}
\label{E-569}
\end{equation}

the Hamiltonian constraint becomes

\begin{equation}
0 = \sqrt{\gamma }\left[ -\sigma G^{\mu \nu \lambda \rho }K_{\mu \nu
}K_{\lambda \rho }-\bar R\right]
= \sqrt{\gamma }\left[ -\sigma \left( K^{\mu \nu }K_{\mu \nu }-K^{2}\right) 
-\bar R\right]
=\mathcal{H}
\label{einstein-3a}
\end{equation}

where we used (\ref{E-527}) and the momentum constraint is

\begin{equation}
0 = -\frac{\partial \mathcal{L}_{\textcolor{dRed}{ADM}}}{\partial N_{\mu }}=
2\sigma \sqrt{\gamma }D_{\nu }\left( G^{\nu \mu \lambda \rho
}K_{\lambda \rho }\right) 
= 2\sigma \sqrt{\gamma } \left( D_{\nu} K^{\nu \mu }- D^\mu K \right) 
=\mathcal{H}^{\mu } \ .
\label{E-571} 
\end{equation}

Comparison with (\ref{einstein-3}) and (\ref{einstein-4}) shows that these
constraints correspond to the non-evolving $G_{\mu 5}$ components of the
Einstein field equations.
Using (\ref{E-532}), we can also write

\begin{equation}
\mathcal{H}^{\nu }=2\sigma \sqrt{\gamma }D_{\mu }\left( G^{\mu \nu
\lambda \rho }K_{\lambda \rho }\right) =\sigma 2D_{\mu }\pi
^{\mu \nu }  \ . 
\label{E-588}
\end{equation}
%


The Legendre transformation to the Hamiltonian density is
\vspace{-12pt}

\begin{eqnarray}
\mathcal{H}_{ADM} \eq \pi ^{\mu \nu }\dot{\gamma}_{\mu \nu }-\mathcal{L}_{\textcolor{dRed}{ADM}}
\left[ \gamma _{\mu \nu },\dot{\gamma}_{\mu \nu },N^{\mu },N\right]
\notag \\
 \eq  -\sigma N\sqrt{\gamma }G^{\mu \nu \lambda \rho }K_{\mu \nu }K_{\lambda
\rho }+2\pi ^{\mu \nu }D_{\mu }N_{\nu }-\sqrt{\gamma }NR
\label{E-589}
\end{eqnarray}

where $N,N^{\mu }$ are Lagrange multipliers and do not require kinetic terms.
Integrating by parts and discarding the total gradient provides

\begin{equation}
2\pi ^{\mu \nu }D_{\mu }N_{\nu }=2D_{\mu }\left( \pi ^{\mu \nu }N_{\nu
}\right) -N_{\nu }\left( 2D_{\mu }\pi ^{\mu \nu }\right) =N_{\nu }\mathcal{H}
^{\nu }  
\label{E-601}
\end{equation}

and using (\ref{einstein-3a}), we arrive at

\begin{equation}
\mathcal{H}_{\textcolor{dRed}{ADM}} = N\mathcal{H}+N_{\nu }\mathcal{H}^{\nu }  \ .
\label{E-606}
\end{equation}

Writing the Hamiltonian in the form

\begin{equation}
\mathcal{H}_{\textcolor{dRed}{ADM}}=\pi ^{\mu \nu }\dot{\gamma}_{\mu \nu }+\dot{p}_{N}\mathcal{
H+}\dot{p}_{N^{\mu }}\mathcal{H}^{\mu }-\mathcal{L}_{\textcolor{dRed}{ADM}}  
\label{E-609}
\end{equation}

the Hamiltonian and momentum constraints $\mathcal{H}=0$ and $\mathcal{H}^{\nu
}=0$ are seen to be 
secondary constraints arising from the requirement that the primary constraints
$p_{N}=0$ and $p_{N^{\mu }}=0$
are preserved under time evolution, 

\begin{equation}
\dot{p}_{N}=\left\{ p_{N},\mathcal{H}_{\textcolor{dRed}{ADM}}\right\} =0\mbox{\qquad}\dot{p}
_{N^{\mu }}=\left\{ p_{N^{\mu }},\mathcal{H}_{\textcolor{dRed}{ADM}}\right\} =0  \ . 
\label{E-610}
\end{equation}

The remaining Einstein equations---the evolution equations $G^{\mu \nu }=0$
--- then follow from

\begin{equation}
\dot{\gamma}_{\mu \nu }=\left\{ \gamma _{\mu \nu },\mathcal{H}_{\textcolor{dRed}{ADM}}\right\} 
\mbox{\qquad}\dot{\pi}^{\mu \nu }=\left\{ \pi ^{\mu \nu },\mathcal{H}
_{\textcolor{dRed}{ADM}}\right\}  
\label{E-611}
\end{equation}

for the canonical variables

\begin{equation}
\left\{ \gamma _{\mu \nu } ,\pi ^{\lambda \rho }
\right\} =\frac{1}{2}\left( \delta _{\mu }^{\lambda }\delta _{\nu
}^{\rho }+\delta _{\mu }^{\rho }\delta _{\nu }^{\lambda }\right)
\label{E-612}
\end{equation}

The equation for $\dot{\gamma}_{\mu \nu }$ reproduces the definition of $\pi
^{\mu \nu }$, since $R$ does not contain $\dot{\gamma}_{\mu \nu }$ and so $
\left\{ \gamma _{\mu \nu },R\right\} =0$. The equation for $\dot{\pi}^{\mu
\nu }$ is thus equivalent to $G^{\mu \nu }=0$.

\section{Perturbations to Schwarzschild Geometry}
\label{SG}

\textcolor{dRed}{To get a feel for some simple possibilities in this formalism,
we pose a Schwarzschild-like interval in an empty pseudo-spacetime $\cm_5$}

\begin{equation}
ds^{2}=-c^{2}Bdt^{2}+Adr^{2}+r^{2}d\theta ^{2}+r^{2}\sin ^{2}\theta d\phi
^{2}+\sigma N^{2}c_{5}^{2}d\tau ^{2}\mbox{\qquad} \qquad T_{\alpha \beta }=0
\label{E-627}
\end{equation}

where $ N = N(x,\tau) $ and we allow the mass parameter $M = M(\tau)$ in the coefficients 

\begin{equation}
B\left( r,\tau \right)  = A^{-1}\left( r,\tau \right) =\left( 1-\frac{
\textcolor{dRed}{2}GM\left( \tau \right) }{rc^{2}}\right)  \qquad \qquad 
N^{2}  = N^{2}\left( t,r,\tau \right)  
\label{E-629}
\end{equation}

to be \textcolor{dRed}{$\tau$}-dependent.  Although the 4D connection and curvature on $\cm$ may
now depend on $\tau$, these~generalizations do not change \textcolor{dRed}{their} structure.
The 4+1 metric is

\begin{equation}
\gamma _{\mu \nu }=\text{diag}\left( -B,A,r^{2},r^{2}\sin ^{2}\theta \right) 
\mbox{\qquad}g_{5\mu } = N_\mu =0\mbox{\qquad}g_{55}=\sigma N^{2}  
\label{E-631}
\end{equation}

and so the Einstein equations reduce to

\begin{equation}
\partial _{\tau }\gamma _{\mu \nu }=-2NK_{\mu \nu }  \qquad \qquad 
\partial _{5}K_{\mu \nu }=-D_{\mu }D_{\nu }N+N\left( -\sigma \bar{R}_{\mu
\nu }+KK_{\mu \nu }-2K_{\mu }^{\lambda }K_{\nu \lambda }\right)
\label{E-645}
\end{equation}

with constraints

\begin{equation}
\bar{R}-\sigma \left( K^{2}-K^{\mu \nu }K_{\mu \nu }\right)  = 0
\qquad \qquad \qquad \qquad 
D_{\mu }K_{\nu }^{\mu }-D_{\nu }K  = 0  
\label{E-648}
\end{equation}

\subsection{Constant Mass Source}

Taking $M\left( \tau \right) =\textcolor{dRed}{m}=\text{constant}$, we find

\begin{equation}
\partial _{5}\gamma _{\mu \nu }=0=-2NK_{\mu \nu } \ \longrightarrow \  K_{\mu \nu
}=0   \ \longrightarrow \  \bar{R}_{\mu \nu } = -\sigma \frac{1}{N
}D_{\mu }D_{\nu }N
\label{E-650}
\end{equation}

for the dynamical equations, as expected for a $\tau$-independent 4D Schwarzschild geometry.  
The~momentum constraint is trivially satisfied, and the Hamiltonian constraint reduces to
$\bar{R}=0$. 
Therefore, we must have

\begin{equation}
\textcolor{dRed}{\bar{R}=\gamma^{\mu \nu }\bar{R}_{\mu \nu }=-\sigma \frac{1}{N}\gamma^{\mu \nu }D_{\mu
}D_{\nu }N=0\longrightarrow \gamma^{\mu \nu }D_{\mu }D_{\nu }N=0  }
\label{E-655}
\end{equation}

meaning that $N$ can be any solution to the source-free 4D wave equation, which
in Schwarzschild geometry is

\begin{equation}
\left[ \partial _{0}^{2}-\frac{B}{r^{2}}\partial _{r}\left( r^{2}B\partial
_{r}\right) \right] N (t,r,\tau) =0 
\label{S-wave}
\end{equation}

where $B$ is given in (\ref{E-629}). Writing the Lagrangian for a test
particle as

\begin{equation}
L =\frac{1}{2}M g_{\alpha \beta} \dot x^\alpha \dot x^\beta 
=\frac{1}{2}M\left[ -c^{2}B\left( r,\tau \right) \dot{t}^{2}+A\left( r,\tau
\right) dr^{2}+r^{2}\dot{\theta}^{2}+r^{2}\sin ^{2}\theta \dot{\phi}
^{2}+\sigma c_{5}^{2}N^{2}\right]
\end{equation}

the equations of motion are 

\begin{eqnarray}
0  \eq \ddot{t}+\frac{\partial _{r}B}{B}\dot{r}\dot{t}+c_{5}
\frac{\partial _{5 }B}{B}\dot{t}+\frac{1}{2}\sigma \frac{c_{5}^{2}}{c^{2}
}\frac{\partial _{t}N^{2}}{B} \\
0  \eq \ddot{r}+\frac{1}{2}\frac{\partial _{r}A}{A}\dot{r}^{2}+\frac{1}{2}c^{2}
\frac{\partial _{r}B}{A}\dot{t}^{2}-\frac{1}{A}r\dot{\theta}^{2}-\frac{1}{A}
r\sin ^{2}\theta \dot{\phi}^{2}+c_{5}\frac{\partial _{5 }A}{A}\dot{r}
-c_{5}^{2}\frac{1}{2}\sigma \frac{\partial _{r}N^{2}}{A} \\
0  \eq r^{2}\ddot{\theta}+2r\dot{r}\dot{\theta}-r^{2}\sin \theta \cos \theta 
\dot{\phi}^{2} \\
0  \eq \ddot{\phi}+2\frac{1}{r}\dot{r}\dot{\phi}+2\cot \theta \dot{\theta}\dot{
\phi}
\end{eqnarray}

which are simplified using the rotational symmetry to put $\theta =\pi /2$.
Writing $\partial _{5 }= (1 / c_5) \partial _{\tau }$, these~become

\begin{eqnarray}
0  \eq \ddot{t}+\frac{\partial _{r}B}{B}\dot{r}\dot{t}+
\frac{1}{2}\sigma \frac{
c_{5}^{2}}{c^{2}}\frac{\partial _{t}N^{2}}{B} \label{t} \\
0  \eq \ddot{r}+\frac{1}{2}\frac{\partial _{r}A}{A}\dot{r}^{2}+\frac{1}{2}c^{2}
\frac{\partial _{r}B}{A}\dot{t}^{2}-\frac{1}{A}r\dot{\phi}^{2}-c_{5}^{2}
\frac{1}{2}\sigma \frac{\partial _{r}N^{2}}{A} \label{r} \\
0  \eq \ddot{\phi}+2\frac{1}{r}\dot{r}\dot{\phi}
\end{eqnarray}

where we used $\partial_\tau B = 0$. The angular equation has the standard solution

\begin{equation}
0 = \ddot{\phi}+2\frac{1}{r}\dot{r}\dot{\phi} \ \longrightarrow \
0=\frac{\ddot{ \phi}}{\dot{\phi}}+2\frac{\dot{r}}{r}
 \ \longrightarrow \  r^{2}\dot{\phi}  = J
\label{J}
\end{equation}

with constant $J$.  In the equation for $t$ we recognize

\begin{equation}
\ddot{t}+\frac{\partial _{r}B}{B}\dot{r}\dot{t}
=\frac{1}{B}\left( B\frac{d\dot{t}}{d\tau }+\frac{dB
}{d\tau }\dot{t}\right) =\frac{1}{B}\frac{d\left( \dot{t}B\right) }{d\tau }
\end{equation}

and so (\ref{t}) becomes

\begin{equation}
0  =  \frac{d\left( \dot{t}B\right) }{d\tau }+\frac{1}{2}\sigma 
\frac{c_{5}^{2}}{c^{2}}\partial _{t}N^{2}
\label{t-2}
\end{equation}

leading to a perturbation in the evolution of the $t$ coordinate, which recovers
the usual relation 

\begin{equation}
\dot{t}  =\textcolor{dRed}{\left( 1-\frac{ 2Gm }{rc^{2}}\right)^{-1}} 
\label{t-dot}
\end{equation}

for $ \partial _{t}N^{2} \rightarrow 0$. 
It is convenient to rewrite (\ref{t-2}) as 

\begin{equation}
\dot{t}  = \frac{1}{B}\left( 1-\sigma \frac{c_{5}^{2}}{c^{2}}\frac{1}{2}
\int^\tau d\tau ~\partial _{t}N^{2}\right) \ .
\label{t-3}
\end{equation}

Using (\ref{J}) and (\ref{t-3}), the radial equation becomes

\begin{equation}
0 = \ddot{r}+\frac{1}{2}\frac{\partial _{r}A}{A}\dot{r}^{2}+\frac{1}{2}c^{2}
\frac{\partial _{r}B}{B}\left( 1-\sigma \frac{c_{5}^{2}}{c^{2}}\frac{1}{2}
\int^\tau d\tau ~\partial _{t}N^{2}\right) ^{2}-\frac{1}{A}\frac{J^{2}}{r^{3}}
-c_{5}^{2}\frac{1}{2}\sigma \frac{\partial _{r}N^{2}}{A}
\end{equation}

which we multiply by $2A\dot{r}$ and use



%
\begin{eqnarray}
\frac{d}{d\tau }\left( A\dot{r}^{2}\right)   \eq \textcolor{dRed}{2A\dot{r}\ddot{r}+
\dot{r}^{3}\partial _{r}A} \\
\frac{d}{d\tau }\left( \frac{J^{2}}{r^{2}}\right)   \eq
\textcolor{dRed}{-2\dot{r}\frac{J^{2}}{r^{3}}} \\
\frac{d}{d\tau }\left( -\frac{1}{B}\right)   \eq
\textcolor{dRed}{\dot{r}\frac{\partial _{r}B}{B^{2}}} \\
\frac{d}{d\tau }N^{2}  \eq \dot{r}\partial _{r}N^{2}+\dot{t}\partial _{t}N^{2}
+\partial_\tau N^{2}
\end{eqnarray}
%

to obtain



%
\begin{equation}
0 = \frac{d}{d\tau }\left[ A\dot{r}^{2}-c^{2}\frac{1}{B}+\frac{J^{2}}{r^{2}}
-c_{5}^{2}\sigma N^{2}\right] +\sigma c_{5}^{2}\left[ \left( \frac{d}{d\tau }
\frac{1}{B}\right) \int^\tau d\tau ~\partial _{t}N^{2}+\frac{1}{B}\partial _{t}N^{2}
+\partial_\tau N^{2} \right] 
\label{r-2}
\end{equation}

where we dropped terms in $c_{5}^{4}$.  Integrating by parts

\begin{equation}
\left( \frac{d}{d\tau }\frac{1}{B}\right) \int^\tau d\tau ~\partial _{t}N^{2}
=\frac{d}{d\tau }\left( \frac{1}{B}
\int^\tau d\tau ~\partial _{t}N^{2}\right) -\frac{1}{B}~\partial _{t}N^{2}
\end{equation}

and so the radial equation (\ref{r-2}) becomes

\begin{equation}
0 = \frac{d}{d\tau }\left[ A\dot{r}^{2}-c^{2}\frac{1}{B}+\frac{J^{2}}{r^{2}}
-c_{5}^{2}\sigma \left( N^{2}-\frac{1}{B}\int^\tau d\tau ~\partial _{t}N^{2}\right) 
\right] +c_{5}^{2}\sigma \partial_\tau N^{2} \ .
\end{equation}

Using (\ref{Ham-1}) the Hamiltonian in these coordinates is

\begin{equation}
K=\frac{1}{2}Mg_{\mu \nu }\dot{x}^{\mu }\dot{x}^{\nu }-\frac{1}{2}
Mc_{5}^{2}g_{55}=\frac{1}{2}M\left( A\dot{r}^{2}-c^{2}\frac{1}{B}+\frac{J^{2}
}{r^{2}}\right) -\frac{1}{2}Mc_{5}^{2}\sigma N^{2}
\end{equation}

and the radial equation is now

\begin{equation}
0 = \frac{d}{d\tau }\left[\textcolor{dRed}{\frac{K}{2M}} 
+c_{5}^{2}\sigma \frac{1}{B}\int^\tau d\tau ~\partial _{t}N^{2} 
\right] +c_{5}^{2}\sigma \partial_\tau N^{2} \ .
\label{ham-f}
\end{equation}

showing that the Hamiltonian, and thus the dynamical mass of the test
particle, is not conserved. 
If, \textcolor{dRed}{for example}, we consider a very short perturbation, so that 

\begin{equation}
N(t,r,\tau) = \alpha (\tau) W(t,r)
\end{equation}

with $\alpha (\tau)$ a narrow distribution centered on $\tau = \tau_0$, then
writing

\begin{equation}
\textcolor{dRed}{\Delta M = \sigma c_{5}^{2}\int d\tau ~\partial _{t}N^{2} \simeq  \sigma
c_{5}^{2}\partial_{t}\textcolor{dRed}{W^2}(t(\tau_0),r(\tau_0))\int d\tau \alpha^2 (\tau)}
\end{equation}

we may integrate (\ref{ham-f}) to obtain

\begin{equation}
\textcolor{dRed}{\frac{K}{2M}} + \Delta M \p \textcolor{dRed}{\left( 1-\frac{ 2Gm
}{rc^{2}}\right)^{-1}} \textcolor{dRed}{\simeq - c_{5}^{2}\sigma   W^2 (t(\tau_0),r(\tau_0))
\left[ \alpha^2 (\infty) - \alpha^2 (-\infty)\right] }
= \kappa = \text{constant} 
\end{equation}

describing a distance-dependent shift in the mass of the test particle.
Thus, while the mass parameter \textcolor{dRed}{$m$} that is associated with a source mass
remains constant, the addition of a $g_{55} $ component to the metric induces
mass transfer in the gravitational field.

\subsection{Variable Mass Source}

As a second example, we put $N=1$ and consider a $\tau$-dependent variation in
the mass $M$ parameter of the metric, as given by 

\begin{equation}
M\left( \tau \right) =m\left[ 1+\alpha \left( \tau \right) \right]
\end{equation}

where the perturbation is small and so 

\begin{equation}
\alpha ^{2} \ll 1 \ \longrightarrow \ 
B = A^{-1} = 1 - \Phi_0  \left[ 1+\alpha \left( \tau \right) \right] 
\end{equation}

\textcolor{dRed}{where by comparison with (\ref{E-629}) we have $\Phi_0 = 2Gm / rc^2$.} 
The 4D connection is now $\tau$-dependent, but retains its unperturbed form
with respect to the coordinates $x^\mu$, so the space remains Ricci flat with $ \ \bar{R} = 0$.
\textcolor{dRed}{We may ask what kind of mass-energy-momentum configuration would
give rise to a Schwarzschild geometry with $\tau$-varying mass parameter, and if
such a configuration can be made consistent with the Hamiltonian and momentum
constraints.}  
The dynamical equation for the metric (neglecting terms in $\alpha^2$ and
$\Phi_0^2$) is

\begin{equation}
\partial _{5}\gamma _{\mu \nu } = -2NK_{\mu \nu } \ \longrightarrow \
K_{\mu \nu }  = -\dfrac{1}{2c_{5}}\partial _{\tau }\gamma _{\mu \nu }
=-\dfrac{\Phi _{0}\dot{\alpha} \left( \tau \right) }{2c_{5}}\text{diag}\left( 1,\dfrac{1}{B^{2}},0,0\right)
\label{K-eqn}
\end{equation}

and raising one index, we find

\begin{equation}
K_{\nu }^{\mu } = \gamma ^{\mu \lambda }K_{\lambda \nu } = -\dfrac{\Phi
_{0}\dot{\alpha}\left( \tau \right) }{ 2c_{5}\textcolor{dRed}{B}}\text{diag}\left( -1,1,0,0\right)
 \ \longrightarrow \ K = K_{\mu }^{\mu } = 0 \ .
\end{equation}

Using $ \bar{R} = 0$, $ N = 1 $, $N^\mu = 0$, $ \left( K_{\mu \nu}
\right)^2 \propto \alpha^2 \approx 0 $, along with (\ref{K-eqn}), the evolution
equation for the extrinsic curvature can be written

\begin{equation}
\dfrac{1}{c_{5}}\partial _{\tau }K_{\mu \nu } = -\dfrac{1}{2c_{5}^{2}}\Phi
_{0}\ddot{\alpha}\left( \tau \right) \text{diag} \left(
1,\dfrac{1}{B^{2}},0,0\right) =\sigma \dfrac{8\pi G}{c^{4}} \left[ S_{\mu \nu
}-\dfrac{1}{2}\gamma _{\mu \nu }\left( S+\sigma \kappa \right) \right] 
\label{k-ev}
\end{equation}

which can be solved for $\alpha (\tau)$ if the energy-momentum tensor is known.  
Because $\bar{R} = K = 0$ and $K^{\mu \nu }K_{\mu \nu } \propto \alpha^2
\Phi_0^2 \approx 0$, we may take the Hamiltonian constraint 

\begin{equation}
\bar{R}-\sigma \left( K^{2}-K^{\mu \nu }K_{\mu \nu }\right)  = -\sigma \frac{
16\pi G}{c^{4}}\kappa
\end{equation}

as the statement that the mass density $\kappa $ is approximately zero.
Thus, the evolution equation (\ref{k-ev}) for $K_{\mu \nu}$ can be satisfied by

\begin{equation}
S_{00} = \textcolor{dRed}{B^2} S_{11} = \left( -\sigma \dfrac{c_{5}^{2}}{c^{2}}\dfrac{16\pi G}{c^{2}}\right)
^{-1}\Phi _{0} \p \ddot{\alpha}\left( \tau \right) \qquad \qquad 
S_{22} = S_{33} = 0 \ \longrightarrow \ S = 0 
\end{equation}

\textcolor{dRed}{describing} a $\tau$-dependent energy density $ S_{00}$ and an
energy-momentum $S_{11} $ flowing into the radial direction.
Using the nonzero Christoffel symbols for the Schwarzschild metric

\begin{equation}
\Gamma _{10}^{0}=\frac{1}{2}\frac{\partial _{r}B}{B}\mbox{\qquad}\Gamma
_{00}^{1}=\frac{1}{2}\frac{\partial _{r}B}{A}  
\label{E-2004}
\end{equation}
\begin{equation}
\Gamma _{11}^{1}=\frac{1}{2}\frac{\partial _{r}A}{A}=-\frac{1}{2}\frac{
\partial _{r}B}{B}\mbox{\qquad}\Gamma _{22}^{1}=-\frac{1}{A}r\mbox{\qquad}
\Gamma _{33}^{1}=-\frac{1}{A}r  
\label{E-2005}
\end{equation}
\begin{equation}
\Gamma _{12}^{2}=\frac{1}{r}\mbox{\qquad}\Gamma _{13}^{3}=\frac{1}{r}
\label{E-2006}
\end{equation}

the momentum constraint 

\begin{equation}
p_{\nu } = D_{\mu }K_{\nu }^{\mu
}-D_{\nu }K = \partial _{\mu }K_{\nu }^{\mu }+K_{\nu }^{\lambda }\Gamma
_{\lambda \mu }^{\mu }-K_{\lambda }^{\mu }\Gamma _{\nu \mu }^{\lambda }
\end{equation}

has components

\begin{eqnarray*}
p_{0} \eq \partial _{r}K_{0}^{1}+K_{0}^{0}\Gamma _{0\mu }^{\mu }-K_{\lambda
}^{\mu }\Gamma _{0\mu }^{\lambda }= K_{0}^{0}\Gamma _{0\mu }^{\mu }-K_{0}^{0}\Gamma
_{00}^{0}-K_{1}^{1}\Gamma _{01}^{1}=0
\\
p_{1}  \eq  \partial _{r}K_{1}^{1}+K_{1}^{\lambda }\Gamma _{\lambda \mu }^{\mu
}-K_{\lambda }^{\mu }\Gamma _{1\mu }^{\lambda } =  -\frac{1}{2}\frac{1}{c_{5}r}\Phi _{0}\dot{%
\alpha}\left( \tau \right) \\
p_{2}  \eq  p_3 = 0
\end{eqnarray*}

which corresponds to a mass current $p_1$ flowing in the radial direction, driving
the varying mass parameter $M(\tau)$ in the metric.
Ascribing $M(\tau)$ to a $\tau $-dependent mass distribution that produces the
energy density $ S_{00}$, energy-momentum $S_{11} $, \textcolor{dRed}{and mass
current $p_{1}$}, 
we see once again that a
variation in source mass will be transferred across spacetime by the induced
gravitational field and, in turn, this field will lead to geodesic motion
corresponding to varying mass in a test event.  

\section{Discussion}

Stueckelberg--Horwitz--Piron (SHP) theory is a
covariant approach to relativistic dynamics developed to address the problem of
time as it arises in electrodynamics.  
In order to account for pair creation/annihilation processes in particular,
and to remove from kinematics any {\em a priori} constraints that may lead to
formal difficulties in describing relativistic interaction in general, SHP poses a theory of
spacetime events $x^\mu $ occurring irreversibly at a chronological time $\tau$. 
By working through the implications of gauge theory at the classical and quantum
levels, SHP introduces five $\tau$-dependent electromagnetic potentials that reduce to
Maxwell fields at $\tau$-equilibrium.  
The equations of SHP electrodynamics suggest a formal 5D symmetry structure
that must be broken to 4+1 representations of O(3,1) on physical grounds.  
The resulting interactions form an integrable system in which event
evolution generates an instantaneous current defined over spacetime at $\tau$,
and, in turn, these currents induce $\tau$-dependent fields that act on other
events at $\tau$. 

In this paper, we extend these ideas into general relativity, posing a 5D
pseudo-spacetime coordinatized by $(x^\mu, \tau)$ and possessing a formal 5D
general diffeomorphism symmetry, which must similarly break to 4+1
representations of geometrical and dynamical symmetries.  
This approach makes SHP general relativity naturally amenable to an unambiguous
4+1 foliation, permitting a $\tau$-dependent generalization of gravitation
that can be decomposed to a set of 4D curvature and matter distribution
structures that evolve in $\tau$.  
We have shown that the 15 Einstein equations in 5D decompose into five constraints on
initial conditions and 10 unconstrained evolution equations for the
gravitational field, equivalent to removing the {\em a priori} constraints from
the 10 Einstein equations in 4D.
\textcolor{dRed}{It is the removal of these constraints that permit mass transfer in SHP
gravitation, just as the absence of a mass-shell constraint permits the exchange
of particles and fields in SHP electrodynamics.}
We~completed the transformation of this system to 
an ADM-like canonical system, although computation is 
generally simpler in the system defined by the intrinsic and
extrinsic curvatures.  
In analogy to SHP electrodynamics, the resulting formulation of general relativity
describes an instantaneous distribution of mass and energy at $\tau$ expressed
through $T_{\alpha\beta} (x, \tau)$, inducing a local metric $g_{\alpha\beta}
(x, \tau)$, which, in turn, determines geodesic equations of motion for any
particular event at~$x^\mu (\tau)$.

As a simple first example of this method, we obtained a nonrelativistic
generalization of Newtonian gravitation in the weak field approximation, by
considering a $\tau$-dependent massive source.  
We saw that the non-constant source mass induces a $\tau$-dependent metric, that, in
turn, leads to geodesic motion for a test event associated with
non-conservation of the Hamiltonian function and, thus, mass variation.
We then considered two generalizations of the Schwarzschild solution.  In the
first, we introduced a non-trivial metric component $g_{55}$ and saw that it
must satisfy a 4D sourceless wave equation.  This generalized plane wave
similarly has the effect of inducing a mass shift in a test event.  
In a second generalization, we treated the mass parameter in the standard
components of the Schwarzschild metric as $\tau$-dependent and solved for the
matter distribution that would produce this perturbation.
We found that the mass density of the matter distribution effectively vanishes,
while the energy density and momentum density into the radial direction drive
the variation of $M$.  
These~mass effects may be compared to Equation (\ref{L-2}), in which we saw that
the SHP electromagnetic field component $f_{5 \alpha } $ permits the exchange of
mass between particles and fields, and is, thus, the condition for
non-conservation of proper time. 
The first term of

\begin{equation}
f_{5 \alpha} = \partial_5 a_\alpha - \partial_\alpha a_5
\end{equation}

induces mass exchange through $\tau$-dependence of the electromagnetic field
$a_\alpha$, in analogy with the $\tau$-dependent gravitational field
$\gamma_{\mu\nu} (x,\tau)$ seen in (\ref{pert-1}) and (\ref{E-629}).
The second term induces mass exchange through a non-trivial fifth field
component $a_5$, in analogy with the $g_{55}$ metric component used in (\ref{S-wave}).

Beyond the theoretical interest in modified gravity, the
4+1 formalism in SHP general relativity offers a potentially significant tool
for the calculation of complex dynamics in numerical relativity.
For~example, the weak field approximation for a single source
event given in Section \ref{weak} may be extended by introducing a second source moving at
nonrelativistic velocity toward the first.
\textcolor{dRed}{Although the equations of motion for a test particle in the resulting field is
not amenable to closed form solution, a straightforward numerical 
solution will take account of the nonlinear evolution of the particle's
coordinate time $t$.  By comparison, a solution using $x^0 = ct$ as the evolution
parameter for the particles and fields will necessarily involve significantly
more computational complexity.}

\textcolor{dRed}{It has been shown \cite{speeds} that Maxwell theory emerges from
SHP electrodynamics by taking $c_5 \rightarrow 0$ and, in this sense, can be seen
as an equilibrium limit as $\tau$-evolution of the field reaches a steady state.
Alternatively, this limit is obtained, under appropriate boundary conditions,
by integration \cite{saad} of the fields and currents
over $\tau$, with the effect of summing at each spacetime point the
contributions from all values of $\tau$.
At equilibrium, the electromagnetic fields become $\tau$-independent and the
fields associated with the potential $a_5$ decouple from matter, so that, while
proper time remains unconstrained, it behaves as a classical conserved quantity.
Thus, by restricting his electromagnetic formalism to the $\tau$-independent
four-vector potential $A^\mu(x)$, Fock remained within standard Maxwell theory.
This~restriction was also applied in the formulations of quantum electrodynamics (QED) by
Schwinger \cite{Schwinger} and Feynman \cite{Feynman-1}, leading to fixed masses
for asymptotic particle states.}

\textcolor{dRed}{As we saw in Equation (\ref{E-480}) the Einstein tensor in SHP
can be split into ten unconstrained components and five constraints among
initial condition.
It remains to be shown that the unconstrained components correspond to the
ten components of the Einstein tensor in 4D, and the five constraints
permit mass exchange but conserve the total mass of matter and fields.
We further expect that, as in electrodynamics, an appropriate restriction in SHP
GR will lead to a decoupling of field components and the conservation of four
of the ten components of the Einstein tensor. 
This restriction should produce a $\tau$-parameterized formulation of standard
4D GR, analogous to Fock's proper-time formulation of Maxwell theory.
While it is evident from (\ref{EL-2}) that taking $c_5$ to zero recovers
the standard spacetime geodesic Equation (\ref{eq-m}), the nonlinearity of the
Einstein field equations makes the problem of extracting a $\tau$-parameterized
field theory considerably more difficult.
These aspects of the theory will be reported in future work.
Such a theory would }
be especially \textcolor{dRed}{useful} in cases of strong gravitational fields. 
We thus expect that calculations of black hole collisions and radiation from
stellar collapse may be improved by posing the initial value problem with
respect to the external evolution parameter $\tau$. 
Numerical calculations of this type are beyond the scope of this paper and
they will be discussed in future~work.

\section{Acknowledgment}It is my pleasure to dedicate this paper to Lawrence P.
Horwitz---teacher, collaborator, and~inspiration---on the occasion of his
90th birthday. 






\section*{References}

%

\bibliographystyle{iopart-num}
\bibliography{all_book_ever}{}
\end{document}